\documentclass[reprint, prb,twocolumn,showkeys,showpacs,superscriptaddress,floatfix]{revtex4-1}
\usepackage{braket}
\usepackage{dsfont}
\usepackage{dutchcal}
\usepackage{amsmath,amssymb,times,color,graphicx,multirow,amsfonts,color}
\usepackage{physics,bm,here}
\usepackage{silence}
\DeclareMathAlphabet\mathcalbm{OMS}{cmsy}{b}{n}
\begin{document}
\title{Geometric orbital magnetization in adiabatic processes}
\author{Luka Trifunovic}
\email{luka.trifunovic@riken.jp}
\affiliation{RIKEN Center for Emergent Matter Science, Wako, Saitama 351-0198, Japan}

\author{Seishiro Ono}
\affiliation{Institute for Solid State Physics, University of Tokyo, Kashiwa 277-8581, Japan}

\author{Haruki Watanabe} 
\email{haruki.watanabe@ap.t.u-tokyo.ac.jp}
\affiliation{Department of Applied Physics, University of Tokyo, Tokyo 113-8656, Japan}
\begin{abstract}
We consider periodic adiabatic processes of gapped many-body spinless
electrons. We find an additional contribution to the orbital magnetization due
to the adiabatic time evolution, dubbed \textit{geometric} orbital
magnetization, which can be expressed as derivative of the many-body Berry
phase with respect to an external magnetic field. For two-dimensional band
insulators, we show that the geometric orbital magnetization generally consists
of two pieces, the topological piece that is expressed as third Chern-Simons
form in $(t,k_x,k_y)$ space, and the non-topological piece that  depends on
Bloch states and energies of both occupied and unoccupied bands.
\end{abstract}
\maketitle
\date\today

\section{Introduction}
We consider periodic adiabatic processes of spinless short-range entangled
phases with period $T$ at zero temperature. The ultimate goal when attacking
this type of time-dependent problems on the general ground would be to obtain
an expression for the physical observables induced by the time evolution in
terms of \textit{instantaneous} eigenstates and eigenenergies of the system.

\begin{figure}
	\begin{center}
		\includegraphics[width=0.6\columnwidth]{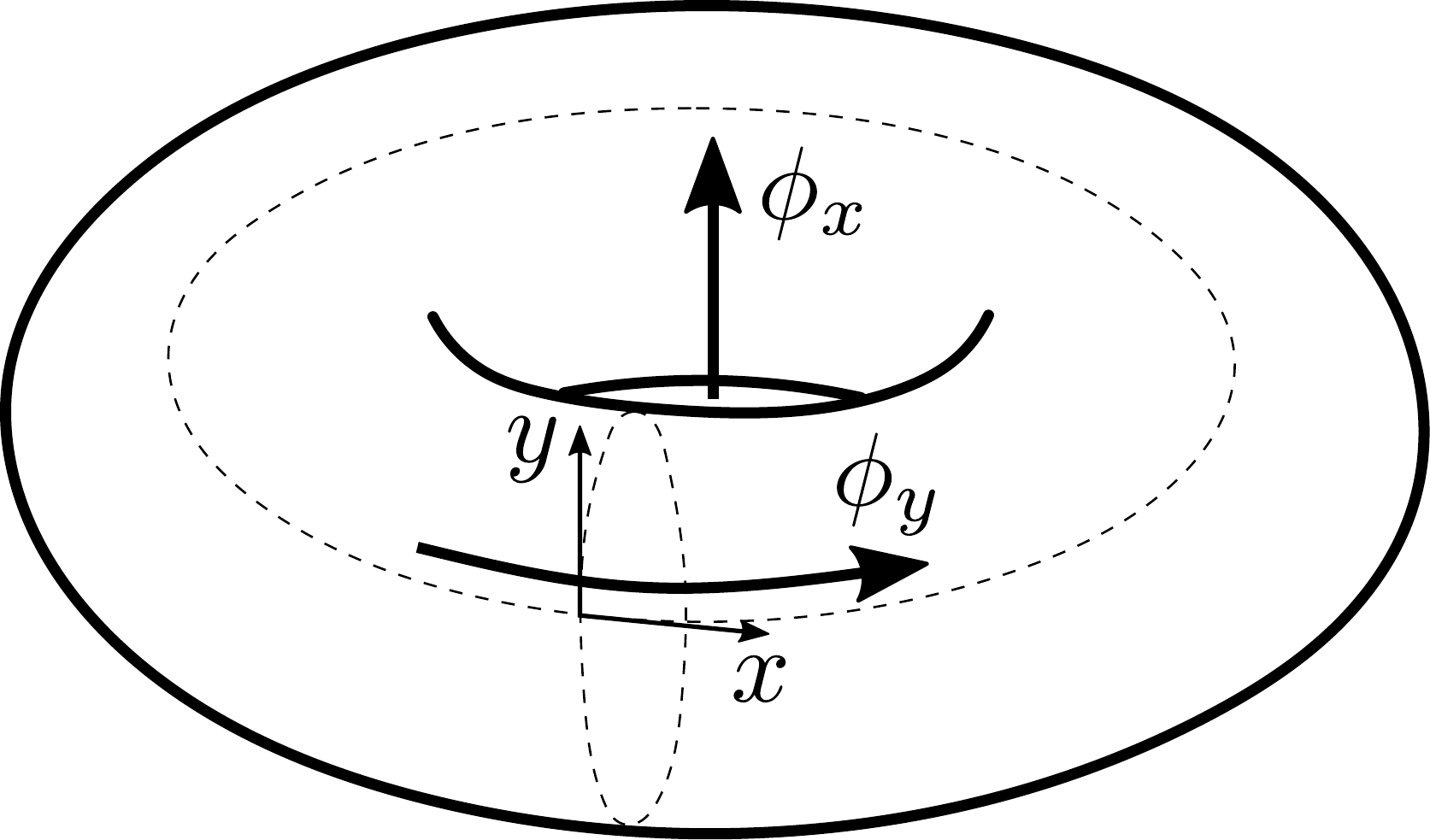}		
		\caption{\label{fig:current} Two dimensional system with
		periodic boundary conditions viewed as toroidal topology. The
		two solenoidal fluxes are denoted by $\phi_x$ and $\phi_y$. 
		The averaged current operator can be expressed as the
	derivative of the Hamiltonian with respect to the fluxes [Eq.~\eqref{eq:currentphi}].}
	\end{center}
\end{figure}

In their pioneering works, Niu and Thouless~\cite{thouless1983,niu1984} found
such an expression for the current operator uniformly averaged over the entire
space. In the formulation, they assumed the periodic boundary conditions with
the period $L_i$ for $i=x,y$ and introduced the solenoidal flux
$\bm{\phi}=(\phi_x,\phi_y)$ as illustrated in Fig.~\ref{fig:current}. 
For concreteness we work in two spatial dimensions throughout this work. Then
the current operator can be expressed as
\begin{equation}
\hat{j}_{t\bm{\phi}}^i\equiv\frac{1}{L_i}\int d^2x\hat{j}_{t\bm{\phi}}^i(\bm{x})=\partial_{\phi_i}\hat{H}_{t\bm{\phi}},\quad i=x,y.\label{eq:currentphi}
\end{equation}
(For brevity we show the dependence on time $t$, flux $\bm{\phi}$, and etc., in
the subscript.) Further taking an average over all values of $\bm{\phi}$, the
expectation value of the current operator induced by the adiabatic
time-evolution can be expressed as the time derivative of the many-body Berry
phase for varying $\bm{\phi}$
\begin{eqnarray}
\int\frac{d^2\phi}{(2\pi)^2}\langle\hat{\bm{j}}_{t\bm{\phi}}\rangle=\partial_t\left(\int\frac{d^2\phi}{(2\pi)^2}\langle\Phi_{t\bm{\phi}}\vert i\partial_{\bm{\phi}}\vert\Phi_{t\bm{\phi}}\rangle\right),	
	\label{eq:jx}
\end{eqnarray}
where $\vert\Phi_{t\bm{\phi}}\rangle$ is the instantaneous ground state of the
Hamiltonian $\hat{H}_{t\bm{\phi}}$.  This expression assumes the periodicity in
$\bm{\phi}$ (see Eqs.~\eqref{eq:p1}, \eqref{eq:p2} below). It is \textit{not}
possible to further impose the periodicity in time simultaneously. Instead we
have $\vert\Phi_{T\bm{\phi}}\rangle
=e^{-i\bm{\phi}\cdot\bm{Q}}\vert\Phi_{0\bm{\phi}}\rangle$, where
$\bm{Q}=\int_0^Tdt\int\frac{d^2\phi}{(2\pi)^2}\langle\hat{\bm{j}}_{t\bm{\phi}}\rangle\in\mathbb{Z}^2$ is the pumped
charge during in one cycle.

The result~(\ref{eq:jx}) is formally similar to the constitutive relation for
Maxwell's equations
\begin{align}
	\bm j(t, \bm x)&=\partial_t\bm p(t, \bm x)+\bm\nabla\times\bm m(t, \bm x),\label{eq:jmeso}
\end{align}
where $\bm p$ and $\bm m$ are the bulk polarization and the bulk magnetization.
Later it was shown~\cite{king-smith1993,vanderbilt1993,resta1994,resta2007}
that Thouless result~(\ref{eq:jx}) combined with constitutive
relation~(\ref{eq:jmeso}) gives a useful formula for the bulk polarization ---
this development marked the birth of ``the modern theory'' of electric
polarization. The bulk polarization is given by the integral of the Berry
connection  (the \textit{first} Chern-Simons form) $P_1$ [see
Eq.~\eqref{eq:modthp} for the formula for band insulators].

Alternatively, let us impose \textit{the periodicity in time}
$\vert\Phi_{T\bm{\phi}}\rangle=\vert\Phi_{0\bm{\phi}}\rangle$ instead.  In this
setting it is useful to integrate over time rather than the solenoidal flux.
Then the Thouless result reads~\cite{niu1984}
\begin{align}
&\int_0^Tdt\langle \hat{\bm{j}}_{t\bm{\phi}}\rangle=-\partial_{\bm{\phi}}\varphi_{\bm{\phi}},
	\label{eq:jTh}\\
&\varphi_{\bm{\phi}}\equiv \int_0^Tdt\langle\Phi_{t\bm{\phi}}\vert i\partial_t\vert\Phi_{t\bm{\phi}}\rangle,
\label{eq:Berry}
\end{align}
where $\varphi_{\bm{\phi}}$ is the many-body Berry phase associated with the
adiabatic time-evolution. In the thermodynamic limit,
$\partial_{\bm{\phi}}\varphi_{\bm{\phi}}$ is independent of
$\bm{\phi}$~\cite{niu1984,PhysRevB.98.155137} and one can set
$\bm{\phi}=\bm{0}$ for instance. There is also a contribution from
$\partial_{\bm{\phi}}E_{t\bm{\phi}}$ in Eqs.~\eqref{eq:jx}, \eqref{eq:jTh} but
it is negligibly small for the same reason. We find this formulation of the
Thouless pump more useful because it can be generalized to wider class of
physical observables as we discuss below.

\begin{figure}
	\begin{center}
		\includegraphics[width=1.0\columnwidth]{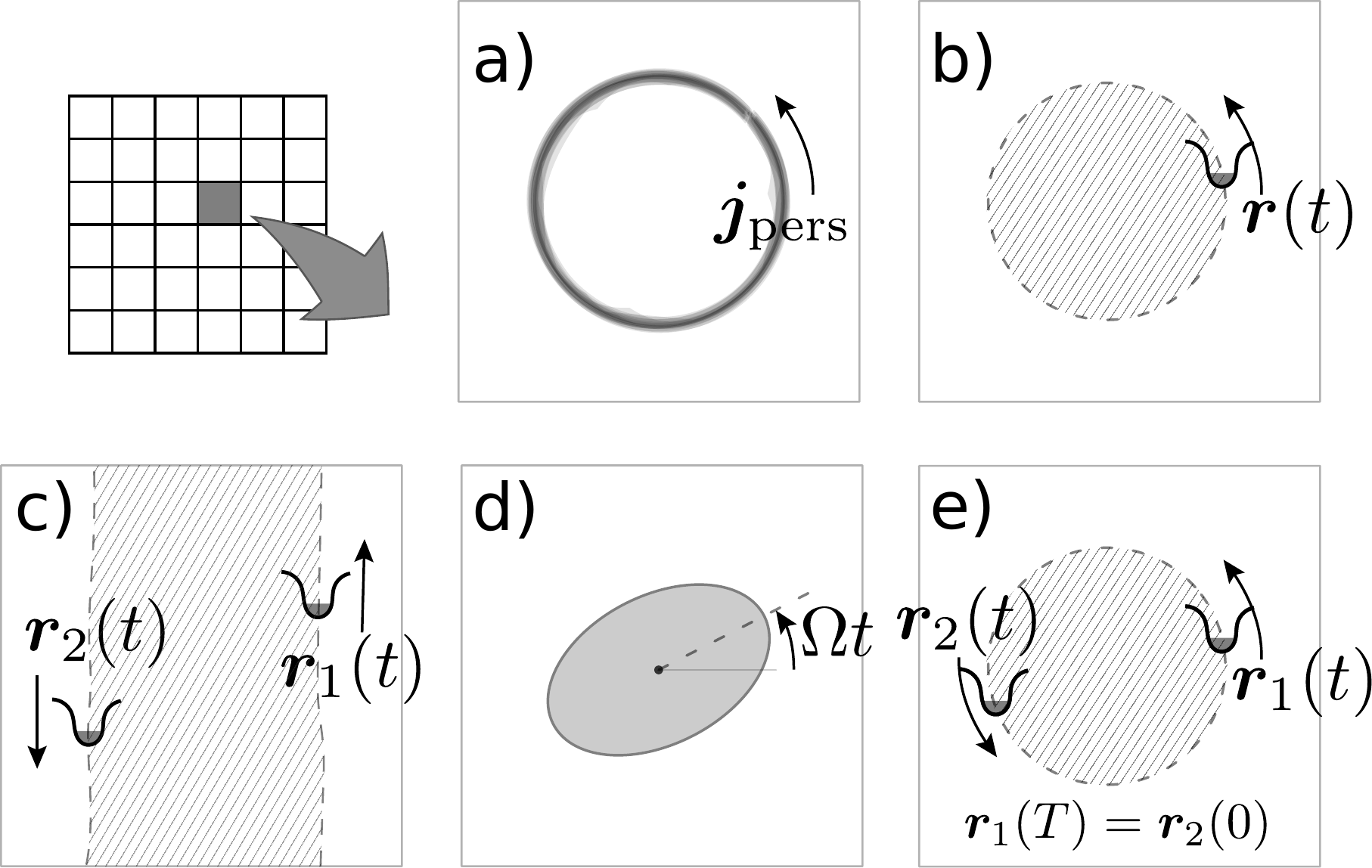}		
		\caption{\label{fig:magnetization}
		Different contributions to orbital magnetization of
		two-dimensional periodic adiabatic process with period $T$.  a): Persistent current $\bm
		j_{\text{pers}}$ within each unit cell produces the
		instantaneous orbital magnetization $\bm m$.  b): An adiabatic
		process where an electron trapped in a potential well whose
		center $\bm{x}=\bm{r}(t)$ is moving along dashed curve. In the
		presence of an externally applied magnetic field, the many-body
		Berry phase $\varphi_{B_z}$ is given by the Aharonov-Bohm flux
		(the hatched area).  c): Periodic boundary conditions are necessary
		when each of two potential wells comes back to its initial
		position after time $T$ by passing though the seam. Two
		possible areas to define Aharonov-Bohm flux (the hatched one and
		the non-hatches one) differ by an integer flux quanta.  d): Unit
		cell consists of a single anisotropic potential well that is
		spinning during adiabatic process. e): Two identical potential
		wells that exchange their positions after single adiabatic
		cycle. All adiabatic processes shown here have vanishing
		integrated current~(\ref{eq:jTh}).}
	\end{center}
\end{figure}

The persistent current associated with a part of the orbital magnetization can also
be expressed using the instantaneous eigenstates and eigenenergies of the
Hamiltonian.  For band insulators, it can be written as the curl of a vector
(see Fig.~\ref{fig:magnetization}a), which together with the constitutive
relation~(\ref{eq:jmeso}), allows one to define the orbital magnetization
$\bm{m}_{\text{pers}}$. (The subscript refers to the contribution associated
with the persistent current.) Alternatively, one can evaluate the change of the
instantaneous ground state energy with respect to the external magnetic field.
This recent development~\cite{xiao2005,thonhauser2005,ceresoli2006,shi2007}
goes under the name of ``the modern theory'' of the orbital magnetization [see
Eq.~\eqref{eq:modthm}]. Unlike the bulk polarization, $\bm{m}_{\text{pers}}$ is
not related to topological response.

In this work we develop a general formulation of the remaining contribution to
the electric current in the constitutive relation ~\eqref{eq:jmeso} that are
neither captured by the averaged current in Eqs.~\eqref{eq:jx}, \eqref{eq:jTh}
nor by the persistent current
$\bm{\nabla}\times\bm{m}_{\text{pers}}(t,\bm{x})$.  We find that, after
coarse-graining in time, this contribution can be expressed as the curl of an
additional term $\mathbcal{m}$ to the orbital magnetization so that $\bm{m}$ in
Eq.~\eqref{eq:jmeso} is given by
\begin{equation}
\bm{m}=\bm{m}_{\text{pers}}+\mathbcal{m}.
\end{equation}
Our main result is that $\mathbcal{m}$ can be obtained as a derivative of the
many-body Berry phase with respect to an external magnetic field $B_z$ applied
in $z$ direction
\begin{align}
	TV\mathcal{m}_z&=\partial_{B_z}\varphi_{B_z}\rvert_{B_z=0},
	\label{eq:calMz}
\end{align}
where $V$ represents the system size and $\varphi_{B_z}$ is defined by
Eq.~(\ref{eq:Berry}) upon substitution $\bm\phi\rightarrow B_z$. This
expression is well-defined in two-dimensional systems with the open boundary
condition at least in one direction. There are known subtleties when applying
uniform magnetic field to periodic systems. See Sec.~\ref{subsec:general} for
the detailed discussion. In the following we assume vanishing Chern numbers in
$(\phi_x,\phi_y)$, $(t,\phi_x)$ and $(t,\phi_y)$ spaces.

As comparison, in the presence of an external uniform magnetic field $\bm
B=(0,0,B_z)^{\rm T}$, the instantaneous orbital magnetization
$\bm{m}_{\text{pers}}$ gives an energy shift
$\int_0^Tdt(E_{tB_z}-E_{t0})/T=-V\bm{m}_{\text{pers}}\cdot\bm B+O(\bm{B}^2)$ of
the many-body ground state. (Throughout this work we set $\hbar=1$.)
Accordingly, after the period $T$, the ground state acquires an additional
phase proportional $TV\bm{m}_{\text{pers}}\cdot\bm B$.  On the other hand, a
non-zero value of $\mathbcal{m}$ shows up as Berry phase
$TV\mathbcal{m}\cdot\bm B+O(\bm{B}^2)$ acquired by the many-body ground state.
For this reason, we name $\mathbcal{m}$ \textit{geometric orbital
magnetization}. The bulk quantity $T\mathbcal{m}$ is independent of the period
$T$ and is defined ``mod $e$''. This ambiguity is reflecting the possibility of
decorating the boundary by one-dimensional Thouless pump.

For band insulators, we perform the perturbation theory with respect to the
applied magnetic field following Refs.~\onlinecite{shi2007,essin2010} and find
that geometric orbital magnetization consists of two contributions 
\begin{equation}
\mathbcal{m}=\mathbcal{m}^{\text{top}}+\mathbcal{m}^{\text{non-top}},
\label{eq:calm2}
\end{equation}
the topological contribution $\mathbcal{m}^{\text{top}}$ is expressed as
integral  of the \textit{third} Chern-Simons form $P_3$ in $(t,k_x,k_y)$ space
[see Eq.~\eqref{eq:3Dtopo2}], while the non-topological contribution
$\mathbcal{m}^{\text{non-top}}$  is written in terms of instantaneous Bloch
states and energies [Eq.~\eqref{eq:nontopcalM}].  The obtained expression for
$\mathbcal{m}$ of band insulators has a formal similarity with the expression
for the magnetoelectric polarizability of three-dimensional band
insulators~\cite{essin2010,Malashevich2010,Chen2011} upon identification
$t/T\leftrightarrow k_z/2\pi$. It is worth mentioning that due to relatively
large gap (order of electronvolts) of band insulators, the adiabaticity
conditions is not particularly restrictive, the period $T$ can be as small as
several femtoseconds.

To gain intuitive understanding of the two contributions~(\ref{eq:calm2}), one
can think of the topological piece $\mathbcal{m}^{\text{top}}$ to be
originating from the Aharanov-Bohm contribution to the many-body Berry phase in
the magnetic field. Thus $\mathbcal{m}^{\text{top}}$ describes the
magnetization from electrons, whose positions are moving during the adiabatic
process, as depicted with dashed lines and arrows in
Fig.~\ref{fig:magnetization}b-c and e. Although Fig.~\ref{fig:magnetization}a
may look similar, $\bm j_{\text{pers}}$ in Fig.~\ref{fig:magnetization}a
represents a \emph{static} persistent current that is uniformly distributed on
the ring.  In contrast, the current density in Fig.~\ref{fig:magnetization}b at
each time is localized to the position of the potential well and it becomes
divergence-free only after averaging over the period $T$. Similarly, the
non-topological piece $\mathbcal{m}^{\text{non-top}}$ can be understood to be
originating from ``spinning'' of anisotropic crystalline potentials, see
Fig~\ref{fig:magnetization}d.

Let us mention at this point several related works. Adiabatic
dynamics can be induced by time-dependent lattice deformations (phonons), which
is the subject of studies on dynamical deformations of
crystals.~\cite{ceresoli2002,juraschek2017,juraschek2018,dong2018,stengel2018}
In Refs.~\onlinecite{juraschek2017,juraschek2018} it was shown that
a time-varying polarization gives rise to a contribution to the orbital
magnetization, and the semi-classical description developed in
Ref.~\onlinecite{dong2018} found the same effect within their framework. The
time-varying polarizations in these works correspond to the situation depicted
in Fig.~\ref{fig:magnetization}b. In the case of band insulators, we find that
they are captured by the \textit{abelian} third Chern-Simons form. Furthermore,
Refs.~\onlinecite{ceresoli2002,stengel2018} showed that rotation of molecules,
as in Fig.~\ref{fig:magnetization}d, gives rise to an orbital magnetization
contribution that can be captured by relation~(\ref{eq:calMz}). The present
approach gives unified description of the above-mentioned effects. More
importantly, it properly describes the orbital magnetization in adiabatic
processes that have not been previously considered: the process in
Fig.~\ref{fig:magnetization}e has inversion symmetry at all times, thus
polarization is time-independent, yet it gives rise to non-zero $\mathbcal{m}$,
which for the case of band insulators is captured by \textit{non-abelian} third
Chern-Simons form, see Sec.~\ref{subsec:tgeoM}.

Analogous to the bulk polarization, crystalline symmetries can quantize topological
geometric orbital magnetization $\mathbcal{m}^{\text{top}}$. We show that,
under certain crystalline symmetries, $\mathbcal{m}^{\text{top}}$ is related to
recently discussed higher-order topological
phases.~\cite{parameswaran2017,schindler2018,peng2017,langbehn2017,song2017,fang2018,ezawa2018,shapourian2017,zhu2018,yan2018,wang2018,wang2018b,khalaf2018,khalaf2018b,trifunovic2019,nobuyuki2018}
Among them, the topological insulators that exhibit quantized corner charges in
the presence of certain crystalline symmetries attracted recently a lot of
theoretical~\cite{benalcazar2017,benalcazar2017,benalcazar2018} and
experimental~\cite{serra-garcia2018,peterson2018} attention. Although, due to
crystalline symmetries, the bulk quadrupole moment is well defined in these
systems (see Fig.~\ref{fig:C4}a), it is still disputed in the literature
whether such definition is possible in the absence of any quantizing crystalline
symmetries.~\cite{kang2018,metthew2018,ono2019} Recent work in
Ref.~\onlinecite{vanmiert2018b} revealed a connection between higher-order
topological
insulators~\cite{parameswaran2017,schindler2018,peng2017,langbehn2017,song2017,fang2018,ezawa2018,shapourian2017,zhu2018,yan2018,wang2018,wang2018b,khalaf2018,khalaf2018b,trifunovic2019,nobuyuki2018}
protected by roto-inversion symmetries and adiabatic processes that involve
topological insulators with quantized corner charges. In
Sec.~\ref{sec:symmetries} we show that the adiabatic processes discussed by van
Miert and Ortix~\cite{vanmiert2018b} are characterized by quantized geometric
orbital magnetization, and we relate the value of $T\mathcal{m}_z^{\text{top}}$ to
the quantized corner charge.

The remaining of this article is organized as follows: in Sec.~\ref{sec:prelim}
we review the modern theory of the polarization and the orbital
magnetization, Sec.~\ref{sec:noninteracting} contains derivation of our main
results, Sec.~\ref{sec:symmetries} discusses the role of symmetries in the
adiabatic process, and Sec.~\ref{sec:examples} presents various non-interacting
examples that illustrate difference between instantaneous orbital
magnetization, topological and non-topological geometric orbital magnetization.
More precisely, we consider toy models illustrating systems depicted in
Fig.~\ref{fig:magnetization}. As a more realistic application of physics
considered in this work, we present in Sec.~{\ref{subsec:rotoM}} calculation of
magnetization induced by rotation of an insulator. A long time
ago,~\cite{barnett1915,barnett1935} Barnett considered magnetization of an
uncharged paramagnetic material when spun on its axis. Modeling paramagnetic
material as collection of local magnetic moments that are randomly oriented,
Barnett~\cite{barnett1915} argued that rotation creates a torque that acts to
align local magnetic moments with rotation axis. This torque gives rise to
magnetization $M=\chi\Omega/\gamma$, where $\chi$ is paramagnetic
susceptibility, $\Omega$ is rotation frequency and $\gamma$ is electron
gyromagnetic ratio. Barnett's measurement of this
effect~\cite{barnett1915} provided first accurate measurement of electron
gyromagnetic ratio. We calculate $\mathbcal{m}$ for this model, which, as seen from
Eq.~(\ref{eq:calMz}), is also proportional to rotational frequency
$\Omega=2\pi/T$ and estimate quantum correction to Barnett effect. Electron
contribution to $\mathbcal{m}$ has both topological and non-topological piece,
but since the system is uncharged, we find that electron contribution to
$\mathbcal{m}^{\text{top}}$ is canceled by corresponding ionic contribution. Thus
resulting $\mathbcal{m}$ is solely due to anisotropy of crystalline potential,
analogous to toy model in Fig.~\ref{fig:magnetization}d. In
Sec.~\ref{sec:examples2} we consider examples of general interacting systems
where periodic adiabatic process consists of
``spinning''~\cite{ceresoli2002,stengel2018} or
``shaking''~\cite{juraschek2017,juraschek2018,dong2018} of the whole
system.~\cite{goldman2014} Our conclusions and outlook can be found in
Sec.~\ref{sec:conclusions}.

\section{Preliminaries}\label{sec:prelim}
Here we review the formulation of the polarization and the orbital
magnetization for band insulators in $2+1$ dimensions developed in
Refs.~\onlinecite{king-smith1993,vanderbilt1993,resta1994,resta2007,xiao2005,thonhauser2005,ceresoli2006,shi2007}.
To simplify notations, we assume primitive lattice vectors of the square
lattice type, but this general framework is \textit{not} restricted to this
special choice. 

\subsection{Modern theory}
Let us denote by
$\psi_{t\bm{k}n}(\bm{x})=(a/\sqrt{V})e^{i\bm{k}\cdot\bm{x}}u_{t\bm{k}n}(\bm{x})$
the instantaneous Bloch function of $n$-th occupied band, satisfying
$h_{t}|\psi_{t\bm{k}n}\rangle=\varepsilon_{t\bm{k}n}|\psi_{t\bm{k}n}\rangle$.
Here $h_t$ is the single-particle Hamiltonian with a periodic potential, $V=L_xL_y$ is the
system size and $a$ is the lattice constant.  We choose the cell-periodic gauge
so that they obey the following conditions for any lattice vector $\bm{R}$ and
reciprocal lattice vector $\bm{G}$~\cite{vanderbilt2018}
\begin{align}
&u_{t\bm{k}n}(\bm{x}+\bm{R})=u_{t\bm{k}n}(\bm{x}),\\
&u_{t\bm{k}+\bm Gn}(\bm{x})=e^{-i\bm  G\cdot\bm{x}}u_{t\bm{k}n}(\bm{x}).\label{eq:uk_b1}
\end{align}

According to the modern theory, the bulk polarization density $\bm{p}(t)$ is
given by
\begin{equation}
\bm{p}(t)=\frac{ei}{V}\sum_{\bm{k}n\in\text{occ}}\langle u_{t\bm{k}n}|\bm{\nabla}_{\bm{k}}u_{t\bm{k}n}\rangle\,\,\text{ mod }\,\,\frac{e}{a}.\label{eq:modthp}
\end{equation}
where $e$ $(<0)$ is the electric charge. The sum over $\bm{k}$ can be replaced
with the integral $V\int \frac{d^2k}{(2\pi)^2}$ over the first Brillouin zone.
Similarly, the orbital magnetization density $\bm{m}_{\text{pers}}(t)$ is given
by
\begin{equation}
\bm{m}_{\text{pers}}(t)=\frac{ei}{2V}\sum_{\bm{k}n\in\text{occ}}\langle\bm{\nabla}_{\bm{k}}u_{t\bm{k}n}|\times(h_{t\bm{k}}+\varepsilon_{t\bm{k}n})|\bm{\nabla}_{\bm{k}}u_{t\bm{k}n}\rangle,\label{eq:modthm}
\end{equation}
where $h_{t\bm{k}}\equiv e^{-i\bm{k}\cdot\bm{x}}h_{t}e^{i\bm{k}\cdot\bm{x}}$. The ambiguity in
Eq.~\eqref{eq:modthp} can be seen by a smooth gauge transformation
$|u_{t\bm{k}n}\rangle'= \sum_m|u_{t\bm{k}m}\rangle(U_{\bm{k}})_{m,n}$ that
changes the integral in Eq.~\eqref{eq:modthp} by an integer multiple of $e/a$,
while the integral in~\eqref{eq:modthm} remains unchanged.

In addition to the derivation via the Thouless pump as we described in the
introduction, the formula \eqref{eq:modthp} was also verified in terms of the
Wannier state localized around the unit cell $\bm{R}$.
\begin{equation}
|w_{tn\bm{R}}\rangle\equiv\frac{a}{\sqrt{V}}\sum_{\bm k}e^{-i\bm k\cdot\bm R}|\psi_{tn\bm k}\rangle.\label{eq:Wannier}
\end{equation}
In terms of the Wannier function, $\bm{p}(t)$ is the deviation of the Wannier center from $\bm{R}$, i.e.,  $\bm{p}(t)=\frac{e}{a^2}\int d^2x(\bm{x}-\bm{R})|w_{t n\bm{R}}(\bm{x})|^2$.

When the origin of unit cell is changed by $\bm{\delta}$, we find
\begin{align}
&|u_{t\bm{k}n}\rangle'=e^{i\bm  k\cdot \bm\delta}|u_{t\bm{k}n}\rangle,\label{origin}\\
&\bm{p}'(t)=\bm{p}(t)-\frac{e\bm{\delta}}{a^2},\\
&\bm{m}_{\text{pers}}'(t)=\bm{m}_{\text{pers}}(t).
\end{align}
Namely, $\bm{p}(t)$ depends on the specific choice of the origin, while $\bm{m}_{\text{pers}}(t)$ does
not. Therefore, it is not $\bm{p}(t)$ itself but rather the change $\Delta \bm{p}(t)$
that is of physical interest. It also follows that for an periodic adiabatic
process, where the system is translated by certain number of unit
cells during the period $T$, the orbital magnetization is periodic in time
while the polarization is not.

For interacting systems under the periodic boundary condition, the combination
in the parenthesis in Eq.~\eqref{eq:jx} replaces Eq.~\eqref{eq:modthp}.  The
periodicity in $\phi_i$ in this formulation is encoded in the relation
\begin{align}
&\vert\Phi_{t2\pi\phi_y}\rangle =e^{-2\pi i\hat{P}_x}\vert\Phi_{t0\phi_y}\rangle,\label{eq:p1}\\
&\vert\Phi_{t\phi_x2\pi}\rangle =e^{-2\pi i\hat{P}_y}\vert\Phi_{t\phi_x0}\rangle,\label{eq:p2}
\end{align}
where $\hat{P}_i$ is the polarization operator (see
Ref.~\onlinecite{watanabe2018} for example). 

\subsection{Topological response}\label{sec:top_res}
To discuss the physical consequence of $\Delta \bm{p}(t)$, let us recall first the
topological linear response in $(1+1)$ dimension~\cite{thouless1983} that holds
at a mesoscopic scale after coarse-graining
\begin{align}
&j^\mu(t,x)=-\sum_\nu\varepsilon^{\mu\nu}\partial_\theta P_1(\theta)\partial_\nu\theta,\label{eq:1Dtopo}\\
&P_1(\theta)\equiv-e\int\frac{dk}{2\pi} \tr A_{\theta k},\label{eq:1Dtopo2}\\
&(A_{\theta k})_{n,m}\equiv-i\langle u_{\theta kn}|\partial_{k}u_{\theta km}\rangle.
\end{align}
Here, $x^\mu$ ($\mu=0,1$) represents $(t,x)$, and $j^\mu$ corresponds to
$(n,j)$. $A_{\theta k}$ is a (finite dimensional) matrix constructed by
\textit{occpied} Bloch states and the trace in Eq.~\eqref{eq:1Dtopo2} is the
matrix trace.  Comparing with above equations, we see that $P_1(\theta(t))$ is
the 1D version of $\bm{p}(t)$ in Eq.~\eqref{eq:modthp}. This response is
derived starting from the Chern-Simons theory
$j^\mu=\frac{C_1}{2\pi}\sum_{\nu\lambda}\varepsilon^{\mu\nu\lambda}\partial_\nu
A_\lambda^{\text{ex}}$ in $(2+1)$ dimensions that describes the response toward
an external field $A^{\text{ex}}$ and reducing the dimension to $(1+1)$
dimensions.

The parameter $\theta$ in Eq.~\eqref{eq:1Dtopo} is a slowly varying field
interpolating between two different systems.  For example, an adiabatic time
evolution $\theta(t)$ induces the bulk current $j(t,x)=\partial_t
P_1(\theta(t))$. The bulk charge transfer from $t=0$ to $t=T$ is thus given
by $\int_{0}^{T}dt j(t,x)=P_1(\theta(T))-P_1(\theta(0))$.
Similarly, a transition of one 1D system to another can be described by
$\theta(x)$, giving rise to a charge density $n(t,x)=-\partial_x
P_1(\theta(x))$. Therefore, the total charge $Q^{\text{edge}}$ accumulated to
the boundary is $Q^{\text{edge}}=\int_{x_0}^{x_1}dx
n(t,x)=P_1(\theta(x_0))-P_1(\theta(x_1))$.  For a given $\theta$ that specifies
$P_1(\theta)$ as a continuous function of $t$ and $x$, even the integer part of
$Q^{\text{edge}}$ is well-defined. However, only the fractional part of
$Q^{\text{edge}}$ is independent of the detailed choice of the interpolation
--- the fractional part depends only on the initial and the final
values of $P_1$ that can be individually computed by Eq.~\eqref{eq:1Dtopo2}.
What we described here can be straightforwardly translated to 2D systems.  The
pumped charge through the bulk per unit length along $\bm n$ is given by
$\bm{Q}\cdot\bm n$, where
\begin{equation}
	\bm{Q}\equiv\int_{0}^{T}dt\,\partial_t\bm{p}(t)=\frac{1}{T}[\bm{p}(T)-\bm{p}(0)].\label{eq:Jb}
\end{equation}

The analog of Eq.~\eqref{eq:1Dtopo} in $(3+1)$ dimensions reads~\cite{qi2008}
\begin{align}
&j^\mu(t,\bm x)=-\frac{1}{2\pi}\sum_{\nu,\lambda,\rho}\varepsilon^{\mu\nu\lambda\rho}\partial_\theta P_3(\theta)\partial_\nu\theta \partial_\lambda A_\rho^{\text{ex}},\label{eq:3Dtopo}\\
&P_3(\theta)\equiv -e\int\frac{d^3k}{8\pi^2}\tr \bm A_{\theta\bm{k}}\cdot(\bm\nabla_{\bm k}+\tfrac{2i}{3}\bm A_{\theta\bm{k}})\times\bm A_{\theta\bm{k}},\\
&(\bm A_{\theta\bm{k}})_{n,m}\equiv-i\langle u_{\theta\bm{k}n}|\bm\nabla_{\bm k}u_{\theta\bm{k}m}\rangle.
\end{align}
Here, $\mu,\nu,\rho,\lambda=0,1,2,3$. Again, $\bm A_{\theta\bm{k}}$ is defined
by occupied Bloch states. This response is derived from the
Chern-Simons theory
$j^\mu=\frac{C_2}{8\pi^2}\varepsilon^{\mu\nu\lambda\rho\sigma}\partial_\nu
A_\lambda^{\text{ex}}\partial_\rho A_\sigma^{\text{ex}}$ in $(4+1)$ dimensions by a
dimensional reduction. This topological response implies, for example, the
magnetoelectric effect~\cite{qi2008,essin2009,essin2010}
$\rho(z)=-\frac{1}{2\pi} \partial_zP_3(\theta(z))B_z^{\text{ex}}$.  

\section{Geometric orbital magnetization}\label{sec:noninteracting}
In this section we present the derivations of our main results.  We start with
verifying the most general expression ~\eqref{eq:calMz} for the geometric
orbital magnetization. Then we derive the formula for the topological and
non-topological contributions in Eq.~\eqref{eq:calm2}.

\subsection{Berry phases in adiabatic process}\label{subsec:general}
Suppose we are interested in the expectation value of the quantity $\hat{X}$,
given by
\begin{equation}
\hat{X}=\partial_\epsilon \hat{H}_{\epsilon}|_{\epsilon=0}\label{eq:other}
\end{equation}
for some parameter $\epsilon$ in the Hamiltonian. For example, in the case of
the averaged current operator, $\epsilon$ can be identified with the solenoidal
flux $\bm{\phi}$ [see Eq.~\eqref{eq:currentphi}]. Likewise, for the orbital
magnetization we use the external magnetic field $B_z$.

Now suppose that the Hamiltonian $\hat{H}_{t}$ has a periodic adiabatic
dependence on $t$, and let $\vert\Phi_{t}\rangle$ be the instantaneous
ground state with the energy eigenvalue $E_{t}$.  We assume an excitation
gap $\Delta_{t}$ and the time-dependence of the Hamiltonian must be slow
enough so that $\Delta_{t}T\gg1$.  Using the density matrix
$\hat\rho_t=\vert\Psi_t\rangle\langle\Psi_t\vert$ obeying the time-dependent
Schr{\"o}dinger equation $\partial_t\hat\rho_t=-i[\hat H_t,\hat\rho_t ]$, we
express the time-average of the expectation value of $\hat{X}_t$ as
\begin{equation}
	X\equiv\int_0^T\frac{dt}{T}{\rm tr}[\hat\rho_t\hat{X}_t].\label{eq:Mz}
\end{equation}
In the absence of the time-evolution, the density matrix is identical to
$\vert\Phi_{t}\rangle\langle\Phi_{t}\vert$.  It acquires contributions from
excited states $\hat{H}_{t}|\Phi_{t}^M\rangle=E_{t}^M|\Phi_{t}^M\rangle$ due to
the time evolution. To the lowest-order perturbation theory with respect to
$(\Delta_{t}T)^{-1}$, the relevant matrix elements are given
by~\cite{thouless1983,watanabe2018}
\begin{equation}
\langle\Phi_{t}^M\vert\hat\rho_t\vert\Phi_{t}\rangle=\langle\Phi_{t}\vert\hat\rho_t\vert\Phi_{t}^M\rangle^*=\frac{i\langle\Phi_{t}^M\vert\partial_t\vert\Phi_{t}\rangle}{E_{t}^M-E_{t}}.
\label{eq:rho_adiabatic}
\end{equation}
Now we plug
$\hat{X}_t\equiv\partial_{\epsilon}\hat{H}_{t\epsilon}|_{\epsilon=0}$ and make
use of the Sternheimer identity
\begin{equation}
(E_{t}-E_{t}^M)\langle\Phi_{t}^M\vert\partial_{\epsilon}\vert\Phi_{t\epsilon}\rangle|_{\epsilon=0}=\langle\Phi_{t}^M\vert\partial_{\epsilon}\hat{H}_{t\epsilon}|_{\epsilon=0}\vert\Phi_{t}\rangle,\label{SI}
\end{equation}
which follows by differentiating
$\hat{H}_{t\epsilon}\vert\Phi_{t\epsilon}\rangle=E_{t\epsilon}\vert\Phi_{t\epsilon}\rangle$
with respect to $\epsilon$ at $\epsilon=0$. In our notation,
$\hat{H}_{t\epsilon}|_{\epsilon=0}=\hat{H}_t$ and
$\vert\Phi_{t\epsilon}\rangle|_{\epsilon=0}=\vert\Phi_{t}\rangle$. Combining
these equations, we find
\begin{align}
	X&=\int_0^T\frac{dt}{T}(\partial_{\epsilon} E_{t\epsilon}+{\cal F}_{t\epsilon})\rvert_{\epsilon=0},\label{eq:BerryF}
\end{align}
where 
\begin{align}
	{\cal F}_{t\epsilon}&\equiv i\partial_t\langle\Phi_{t\epsilon}\vert \partial_{\epsilon}\vert\Phi_{t\epsilon}\rangle-i\partial_\epsilon\langle\Phi_{t\epsilon}\vert \partial_t\vert\Phi_{t\epsilon}\rangle,
\end{align}
is the Berry curvature in $(t,\epsilon)$ space. Further assuming the
periodicity in time
\begin{align}
	\vert\Phi_{T\epsilon}\rangle=\vert\Phi_{0\epsilon}\rangle,\label{eq:GSp}
\end{align}
we arrive at our general expression
\begin{align}
	X&=X_{\text{inst}}+X_{\text{geom}},\\
	X_{\text{inst}}&\equiv\int_0^T\frac{dt}{T}\langle\Phi_t\vert\hat{X}_t\vert\Phi_t\rangle=\int_0^T\frac{dt}{T}\partial_{\epsilon} E_{t\epsilon}\rvert_{\epsilon=0},\label{eq:Xinst}\\
	X_{\text{geom}}&\equiv -\frac{1}{T}\partial_\epsilon\varphi_\epsilon\rvert_{\epsilon=0},\quad\varphi_\epsilon\equiv \int_0^Tdt\langle\Phi_{t\epsilon}\vert i\partial_t\vert\Phi_{t\epsilon}\rangle,\label{eq:Xgeom}
\end{align}
where $X_{\text{inst}}$ is the time average of the expectation value using the
instantaneous ground state and $X_{\text{geom}}$ is the geometric contribution
originating from the adiabatic time dependence.  This is the generalization of
Eq.~\eqref{eq:jTh} for the electric current to physical observables written as
the derivative of Hamiltonian as in Eq.~\eqref{eq:other}. The following
basis-independent expressions may also be useful 
\begin{align}
	X_{\text{geom}}&=\int_0^T\frac{dt}{T} i{\rm tr}\hat P_{t\epsilon}[\partial_t\hat P_{t\epsilon},\partial_\epsilon\hat P_{t\epsilon} ]|_{\epsilon=0}\label{eq:MzSPint}\\
	&={\rm Re}\int_0^T\frac{dt}{T}\oint\frac{dz}{\pi}{\rm tr}[(\partial_t\hat{H}_{t})\hat{G}_{t}^2\partial_{\epsilon}\hat{H}_{t\epsilon}|_{\epsilon=0}\hat{G}_{t}],
\end{align}
where $\hat P_{t\epsilon}=|\Phi_{t\epsilon}\rangle\langle\Phi_{t\epsilon}\vert$
is the projector onto the many-body ground state,
$\hat{G}_{t}=(z-\hat{H}_{t})^{-1}$ is the many-body Green function, and the
integration contour encloses only the ground state at $z=E_{t}$.

Let us now specialize to the case $\epsilon=B_z$. Then $X_{\text{geom}}$ in
Eq.~\eqref{eq:Xgeom} gives the geometric orbital magnetization $\mathbcal{m}$
in Eq.~\eqref{eq:calMz}, while  $X_{\text{inst}}$ is the persistent
current contribution. Note the additional minus sign because of
$\hat{M}_z=-\partial_{B_z} \hat{H}_{B_z}$. Previously,
Refs.~\onlinecite{ceresoli2002,stengel2018} considered the Berry curvature in
$(t,B_z)$ space to describe orbital magnetization induced by rotation of
molecules. See also examples in Sec.~\ref{subsec:nontgeoM}
and~\ref{subsec:rotom} below.

When applying these formulae, one has to be careful about boundary conditions.
If open boundary conditions in at least one direction are imposed, the
result~(\ref{eq:MzSPint}) is directly applicable. However, a process that is
periodic in time under periodic boundary condition may loose its
periodicity in time under open boundary conditions. For example, the system in
Fig.~\ref{fig:magnetization}c is not periodic in time if the open boundary
condition in $y$ direction is imposed. Similarly, the periodicity in time
requires periodic boundary conditions in both directions for the $C_4$-symmetric
system in Fig.~\ref{fig:C4e}. Keeping (original) periodic boundary conditions
in both directions in the presence of magnetic field, implies that the net flux
through the system has to vanish. If the system is homogeneous, local
contributions to $\mathbcal{m}$ cancel out and we cannot obtain a useful
information about the system. (For single-particle problems there is a
resolution as we discuss below.) Finally, one can impose \textit{magnetic}
periodic boundary conditions assuming that the total magnetic flux applied to
the system $B_zL_xL_y$ is an integer multiple of
$2\pi$.~\cite{brown1964,zak1964} However, each eigenstate of $\hat{H}_{B_z}$
may not be analytic as a function of $B_z$ despite the fact that the magnetic
field $B_z=2\pi/L_xL_y$ itself can be made small for a large systems. The
expression~\eqref{eq:MzSPint} is still applicable if the projector onto the
instantaneous ground state is analytic function of $B_z$, which is the case for
band insulators with vanishing Chern
number.~\cite{essin2010,Malashevich2010,gonze2011,Chen2011} However, to our
knowledge there is no general proof for gapped interacting systems. 

\subsection{Noninteracting systems}
\label{nis}
Let us apply this general expression to noninteracting
electrons described by the quadratic Hamiltonian
\begin{equation}
\hat{H}_{t\epsilon}=\sum_{n}\varepsilon_{t\epsilon n}\hat{\gamma}_{t\epsilon n}^\dagger \hat{\gamma}_{t\epsilon n}.
\end{equation}
We label single-particle states in such a way that $\varepsilon_{t\epsilon
n+1}\geq\varepsilon_{t\epsilon n}$ for all $n=1,2,\cdots$. We also assume a
finite gap $\Delta=\varepsilon_{t\epsilon N+1}-\varepsilon_{t\epsilon N}$
between $N$-th and $(N+1)$-th levels. We write the single particle state
$|\gamma_{t\epsilon n}\rangle\equiv \hat{\gamma}_{t\epsilon
n}^\dagger|0\rangle$.  Then the $N$-particle ground state can be written as
\begin{equation}
|\hat{\Phi}_{t\epsilon}\rangle=\prod_{n=1}^N\hat{\gamma}_{t\epsilon n}^\dagger |0\rangle.
\end{equation}
For a later purpose, we allow for a unitary transformation \textit{among the
occupied levels}
\begin{equation}
\hat{\psi}_{t\epsilon \ell}^\dagger=\sum_{n=1}^N\hat{\gamma}_{t\epsilon n}^\dagger U_{n\ell},\quad \ell=1,2,\cdots,N.
\end{equation}
Although such a basis change may sound unnecessary, in the actual application
of this framework it is sometimes important to work in the proper basis by
choosing $U_{n\ell}$ appropriately. (See Sec.~\ref{subsec:GOMBI} for an
example).  After all we find that the many-body Berry phase
$\varphi_{\epsilon}$ is given by the sum of single-particle Berry phases
$\varphi_{\epsilon\ell}$ of occupied levels
\begin{align}
\varphi_{\epsilon}= \sum_{\ell=1}^N\varphi_{\epsilon\ell},\label{eq:manysingle}\quad\varphi_{\epsilon\ell}\equiv\int_0^Tdt\langle\psi_{t\epsilon\ell}\vert i\partial_t\psi_{t\epsilon\ell}\rangle.
\end{align}
Therefore, we get the following expressions for single-particle problems. The
latter two expressions are basis-independent
\begin{align}
X_{\text{geom}}
&=\int_0^T\frac{dt}{T}\,i\sum_{\ell=1}^N\langle \partial_t\psi_{t\epsilon\ell}\vert \partial_\epsilon\psi_{t\epsilon\ell}\rangle|_{\epsilon=0}\\
&=\int_0^T\frac{dt}{T}\,i{\rm tr}P_{t\epsilon}[\partial_t P_{t\epsilon},\partial_\epsilon P_{t\epsilon}]|_{\epsilon=0}\\
&={\rm Re}\oint\frac{dz}{\pi}\int_0^T\frac{dt}{T}\tr[(\partial_th_{t})g_{t}^2\partial_{\epsilon}h_{t\epsilon}\rvert_{\epsilon=0}g_{t}],
	\label{eq:MzSP}
\end{align}
where
$P_{t\epsilon}=\sum_{\ell=1}^N\vert\psi_{t\epsilon\ell}\rangle\langle\psi_{t\epsilon\ell}\vert=\sum_{\ell=1}^N\vert\gamma_{t\epsilon
n}\rangle\langle\gamma_{t\epsilon n}\vert$ is the projector onto occupied
single-particle states, $h_{t\epsilon}=\sum_n\varepsilon_{t
n\epsilon}\vert\gamma_{t \epsilon n}\rangle\langle\gamma_{t \epsilon n}\vert$
is the single-particle Hamiltonian, $g_{t}=(z-h_{t})^{-1}$ is single-particle
Green function, and the integration contour encloses all the occupied states at
$z=\varepsilon_{t n}$ ($n=1,2,\cdots,N$).

For the orbital magnetization, we again set $\epsilon=B_z$. The
same remarks as in the previous section apply here. In the case of band
insulators, one may want to impose periodic boundary conditions to preserve the
translation symmetry. As discussed in the previous section there are two
possibilities to achieve this. One can change the the boundary condition to
magnetic periodic boundary conditions.~\cite{brown1964,zak1964} For
single-particle systems, assuming symmetric gauge $\bm A^{\rm ex}(\bm x)=\bm
B\times\bm x/2$, the magnetic periodic boundary conditions can be taken into
account explicitly by restricting the form of the projector $\hat{P}_{tB_z}$
to~\cite{essin2010,gonze2011} $\langle\bm x_1\vert
\hat{P}_{tB_z}\vert\bm x_2\rangle=\hat{P}^\prime_{tB_z}(\bm x_2,\bm x_1)
e^{ieB_z\bm x_1\times\bm x_2\cdot\hat{\bm z}/2}$, where $P^\prime_{tB_z}(\bm
x_1,\bm x_2)$ is an arbitrary $N\times N$ matrix function (not necessarily
projector) that satisfies $P^\prime_{tB_z}(\bm x_1+\bm R,\bm x_2+\bm
R)=P^\prime_{tB_z}(\bm x_1,\bm x_2)$, where $\bm R$ is an element of Bravais
lattice.  The expression for $\hat{P}^\prime_{tB_z}(\bm x)$ can be found
perturbatively in $B_z$,~\cite{essin2010,gonze2011} which, after substituting
back to Eq.~(\ref{eq:manysingle}), yields an expression for the Berry phase and
$\mathbcal{m}$. The second option is to apply a spatially modulating magnetic
field as we discuss below. 

\subsection{Geometric orbital magnetization for band insulators}
\label{subsec:GOMBI}
Below we consider band insulators and show that geometric orbital magnetization
has two contributions as in Eq.~\eqref{eq:calm2}. Since we assume the periodic
boundary condition both in $x$ and $y$, we apply a slowly modulating magnetic
field~\cite{shi2007,essin2010} in order to avoid changing of the boundary
condition as discussed in the previous subsection. We use the vector potential
\begin{align}
&\bm{A}^{\text{ex}}(\bm{x})=\frac{\epsilon}{2q}(-\sin qy,\sin qx,0)^{\rm T},\\
&\bm{B}(\bm{x})=\bm{\nabla}\times\bm{A}^{\text{ex}}(\bm{x})=\bm{e}_z\epsilon f(\bm{x})
\end{align}
with $\bm{e}_z\equiv(0,0,1)^{\rm T}$, $q\equiv2\pi/L$, and $f(\bm{x})=(\cos
qx+\cos qy)/2$. (To simplify the notation, we assume $L=L_x=L_y$ in this
subsection). Such a magnetic field induces the change of the Bloch function
\begin{align}
&|\partial_\epsilon\psi_{t\epsilon n\bm{k}}\rangle|_{\epsilon=0}\notag\\
&=-\sum_{n'\bm{k}'}|\psi_{tn'\bm{k}'}\rangle\frac{\langle\psi_{tn'\bm{k}'}|\partial_\epsilon h_{t\epsilon}|_{\epsilon=0}|\psi_{tn\bm{k}}\rangle}{\varepsilon_{tn'\bm{k}'}-\varepsilon_{tn\bm{k}}}
\end{align}
and $|\partial_\epsilon w_{t\epsilon n\bm{R}}\rangle|_{\epsilon=0}$ is given
via Eq.~\eqref{eq:Wannier}.

We compute the Berry phase using the formula~\eqref{eq:manysingle} derived
above. It is important to work in the Wannier basis for which the magnetic
field effectively becomes uniform in the limit $q\rightarrow0$. The
single-particle Berry phase in this basis, summed over occupied bands, takes
the following form
\begin{align}
	\partial_{\epsilon}\varphi_{\epsilon \bm{R}}|_{\epsilon=0}&=-\int_0^Tdt\sum_{n\in\text{occ}}\langle i\partial_tw_{tn \bm{R}}\vert \partial_{\epsilon}w_{t\epsilon n \bm{R}}\rangle|_{\epsilon=0}+\text{c.c.}\notag\\
	&=Ta^2\mathcal{m}_zf(\bm{R}).
\end{align}
If we further sum over $\bm{R}$, or equivalently if we work in the Bloch basis
$|\psi_{tn\bm{k}}\rangle$, we get $0$ reflecting the fact that for bulk systems
Fourier component $\mathcal{m}_z(\bm q)$ vanishes for $\bm q\neq0$. Thus, care
must be taken to correctly read off local contribution to $\mathcal{m}_z$--- an
unintentional integration over $\bm R$ of a term proportional to $f(\bm R)$
makes it impossible to find the correct value of $\mathcal{m}_z$. The rest
calculation follows the appendix in Ref.~\onlinecite{essin2010}. Upon taking
the limit $q\rightarrow0$, we find
\begin{widetext}
\begin{align}
\mathcal{m}_z&=\lim_{q\rightarrow0}\frac{1}{T}\int_0^Tdt\int\frac{d^2k}{(2\pi)^2}\sum_{n\in\text{occ}}\sum_{n'\bm{k}'}\,\langle i\partial_t\psi_{tn\bm{k}}|\psi_{tn'\bm{k}}\rangle \frac{\langle\psi_{tn'\bm{k}}|\partial_\epsilon h_{t\epsilon}|_{\epsilon=0}|\psi_{n\bm{k}'}\rangle}{\varepsilon_{tn'\bm{k}}-\varepsilon_{tn\bm{k}'}}+\text{c.c}.\notag\\
&=-\frac{e}{2T}\int_0^Tdt\int\frac{d^2k}{(2\pi)^2}\sum_{n\in\text{occ}}\sum_{n'}\langle\partial_t u_{n\bm k}\vert u_{n^\prime\bm k}\rangle
\frac{\langle u_{tn'\bm k}\vert\bm\nabla_{\bm k}(h_{t\bm k}+\varepsilon_{tn\bm k})\times\vert\bm\nabla_{\bm k}u_{tn\bm k}\rangle}{\varepsilon_{tn'\bm k}-\varepsilon_{tn\bm k}}+\text{c.c}.	
\end{align}
This last expression can precisely be expressed as the sum of two terms,
$\mathbcal{m}^{\text{top}}+\mathbcal{m}^{\text{non-top}}$. The topological
piece $\mathbcal{m}^{\text{top}}$ reads
\begin{align}
&\mathbcal{m}^{\text{top}}= \bm{e}_zP_3/T,\label{eq:topc}\\
&P_3\equiv -\frac{e}{2}\int_0^Tdt\int\frac{d^2k}{(2\pi)^2}\tr[\bm A_{\bm{K}}\cdot\bm\nabla_{\bm{K}}\times\bm A_{\bm{K}}+\tfrac{2i}{3}\bm A_{\bm{K}}\cdot\bm A_{\bm{K}}\times\bm A_{\bm{K}}].\label{eq:3Dtopo2}
\end{align}
The Berry connection $(\bm A_{\bm{K}})_{n,m}\equiv-i\langle
u_{\bm{K}n}|\bm\nabla_{\bm{K}}u_{\bm{K}m}\rangle$ is defined using occupied
Bloch states as a function of $\bm{K}\equiv(t,\bm{k})$. The smoothness and the
periodicity of $\bm A_{\bm{K}}$ are assumed in the integral in
Eq.~\eqref{eq:3Dtopo2}. Such a choice is possible only  when both the pumped
charge through the bulk $\bm Q$ in Eq.~\eqref{eq:Jb} and the 2D Chern
number for $(k_x,k_y)$ vanish.  

The non-topological contribution depends also on instantaneous eigenenergies of the Bloch Hamiltonian
\begin{align}
	\mathbcal{m}^{\text{non-top}}&=\sum_{n\in\text{occ}}\sum_{n'\in\text{unocc}}\frac{e}{2T}\int_0^Tdt\int\frac{d^2k}{(2\pi)^2}\frac{
	\langle u_{tn\bm k}|\partial_tP_{t\bm k}|u_{tn'\bm k}\rangle\langle u_{tn'\bm k}|\{\bm\nabla_{\bm k}h_{t\bm k}\times\bm\nabla_{\bm k}P_{t\bm k}\}|u_{tn\bm k}\rangle
	}{\varepsilon_{tn\bm k}-\varepsilon_{tn'\bm k}}+\text{c.c.}\notag\\
	&=\frac{e}{2T}\int_0^Tdt\int\frac{d^2k}{(2\pi)^2}\oint\frac{dz}{2\pi i}{\rm tr}\left[ \partial_tP_{t\bm k}g_{t\bm k}\{\bm\nabla_{\bm k}h_{t\bm k}\times\bm\nabla_{\bm k}P_{t\bm k}\}g_{t\bm k} \right]+\text{c.c.}
	\label{eq:nontopcalM}
\end{align}
\end{widetext}
Here, $h_{t\bm k}$ is Bloch Hamiltonian, $P_{t\bm k}=\sum_{n\in\text{occ}}\vert u_{tn\bm{k}}\rangle\langle u_{tn\bm{k}}\vert$ is the projector
onto occupied bands at $\bm k$, $g_{t\bm k}=(z-h_{t\bm k})^{-1}$ is Bloch's
Green function, the curly brackets denote symmetrization $\{\bm A\times\bm
B\}=\bm A\times\bm B+\bm B\times\bm A$, and the integration contour encloses
all the filled Bloch states at $z=\varepsilon_{tn\bm{k}}$. See the appendix of Ref.~\onlinecite{essin2010}
for the details. Note that both $\mathbcal{m}^{\text{top}}$ and $\mathbcal{m}^{\text{non-top}}$ are not affected by the shift of the origin in Eq.~\eqref{origin}.

\subsection{Topological contribution from response theory}\label{subsec:topom}
\begin{figure}
	\begin{center}
		\includegraphics[width=0.7\columnwidth]{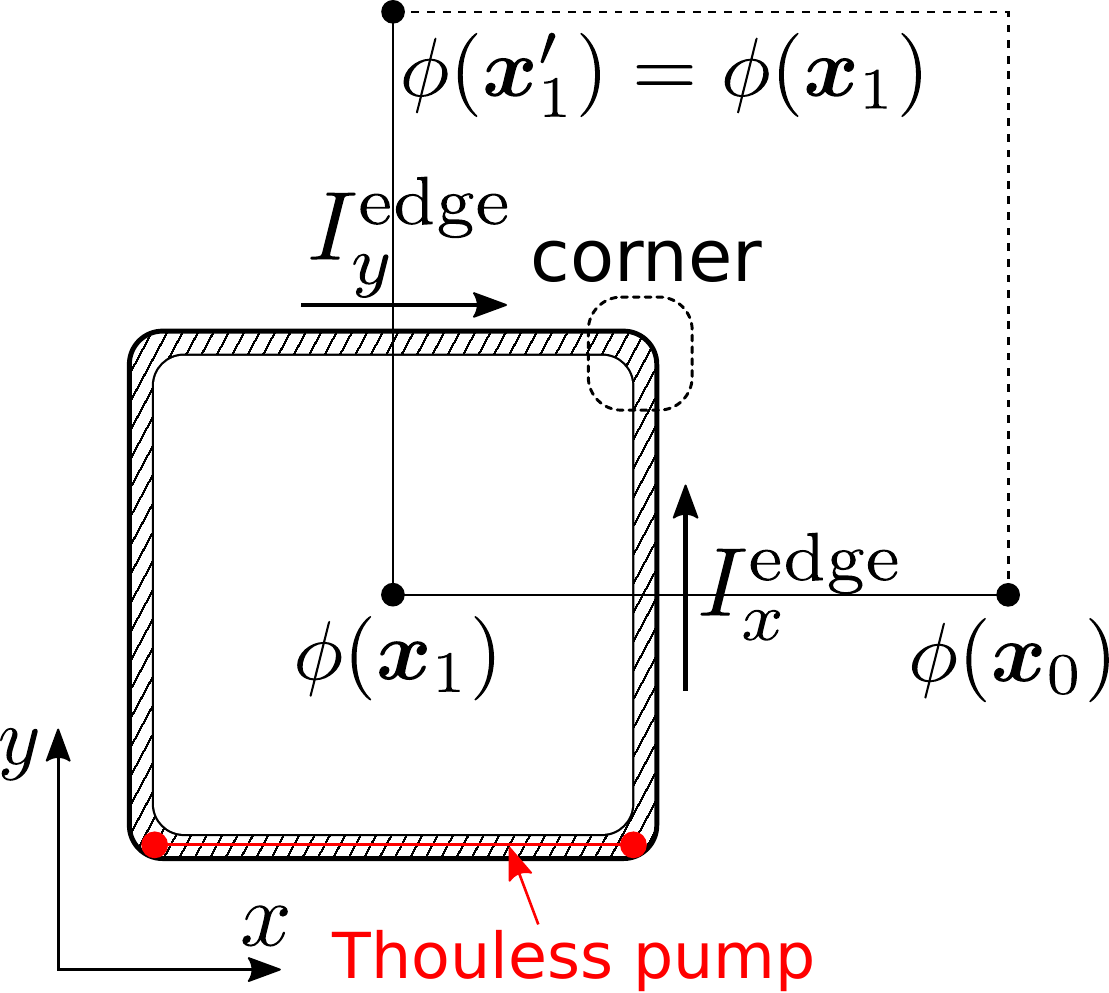}		
		\caption{\label{fig:Qboundary}The boundary current along the
		interface of two adiabatic processes $h_{\phi(\bm x_i)\theta(t)\bm{k}}$ with $i=0$ and $1$. A 1D decoration with Thouless
		pump changes the edge charge transfer by an integer and leads
		to integer corner charge accumulation. Hatched parts denote the boundary area between the two systems.}
	\end{center}
\end{figure}

Here we give an alternative, easier derivation of $\mathbcal{m}^{\text{top}}$
in Eq.~\eqref{eq:topc} from the topological response theory.  To this end let
us further reduce one spatial dimension in Eq.~\eqref{eq:3Dtopo} to achieve the
topological quadratic response in $(2+1)$d~\cite{qi2008}:
\begin{align}
&\mathcal{j}^\mu(t,\bm x)=-\frac{1}{2\pi}\sum_{\nu,\lambda,\rho}\varepsilon^{\mu\nu\lambda}G_2(\theta,\phi)\partial_\nu\theta \partial_\lambda \phi,\label{eq:2Dtopo}\\
&\frac{1}{2\pi}G_2(\theta,\phi)\equiv-e\int\frac{d^2k}{32\pi^2} \,\varepsilon^{\mu\nu\rho\sigma}\tr F_{\mu\nu} F_{\rho\sigma},\label{eq:2Dtopo2}
\end{align}
where $F_{\mu\nu}\equiv\partial_\mu A_{\nu}-\partial_\nu
A_{\mu}+i[A_{\mu},A_{\nu}]$ is the Berry curvature in the
$(k_x,k_y,\theta,\phi)$ space and $\theta$ and $\phi$ are two slowly varying
fields:  $\theta(t)$ denotes an adiabatic and periodic time dependence and
$\phi(\bm{x})$ describes a smooth interface of domains
(Fig.~\ref{fig:Qboundary}). In this setting, we find
\begin{align}
\mathbcal{j}(t,\bm{x})=\frac{1}{2\pi}G_2(\theta,\phi)\partial_t\theta(t)\bm{\nabla}\phi(\bm{x})\times\bm{e}_z
\end{align}
so that
\begin{align}
\label{eq:j4}
	{\mathbcal{j}}(\bm{x})&\equiv\int_{0}^{T}\frac{dt}{T}\mathbcal{j}(t,\bm{x})=\partial_\phi P_3(\phi)\bm{\nabla}\phi(\bm{x})\times\bm{e}_z/T\notag\\
	&=\bm{\nabla}\times[\bm{e}_zP_3(\phi(\bm{x}))/T]=\bm{\nabla}\times\mathbcal{m}^{\text{top}}(\bm{x}).
\end{align}
This reproduces Eq.~\eqref{eq:topc}.  In the derivation we used the relation $\int_0^{2\pi}\frac{d\theta}{2\pi}
G_2(\theta,\phi)=\partial_\phi P_3(\phi)$.  It is important to note that
$\mathbcal{j}(t,\bm x)$ itself cannot be written as a curl of a vector field ---
Eq.~\eqref{eq:j4} holds only after the time convolution (or equivalently the
time average). 

The above derivation relies on the connection~(\ref{eq:jmeso}) between
$\mathbcal{m}^{\text{top}}$ and topological edge current in adiabatically
driven two-dimensional systems. To see this more concretely, let us consider
the boundary of two regions with $\phi_0\equiv\phi(\bm x_0)$ and
$\phi_1\equiv\phi(\bm x_1)$ (see Fig.~\ref{fig:Qboundary}). Just like in the
case of polarization, only the fractional part of the edge current is the bulk
contribution that depends only on $\phi_0$ and $\phi_1$. This can be understood
by noticing that decorating the boundary with a 1D chain leads to an integer
charge transfer through the Thouless pump.~\cite{thouless1983} To capture the
fractional bulk contribution to the edge current, one can separately compute
$\mathcal{m}_z(\bm x_0)$ and $\mathcal{m}_z(\bm x_1)$ without paying attention
to their continuity. The geometric contribution to the charge transfer along
$i$ direction, $i=x,y$ between two bulk systems with $\mathcal{m}_z(\bm{x}_0)$
and $\mathcal{m}_z(\bm{x}_1)=\mathcal{m}_z(\bm{x}_1')$
(Fig.~\ref{fig:Qboundary}) is given by
\begin{align}
I_i^{\text{edge}}&\equiv\int_{\bm  x_0}^{\bm x_1} dx\mathcal{j}_i(\bm{x})=\mathcal{m}_z(\bm x_0)-\mathcal{m}_z(\bm x_1')\mod e.
\end{align}

Notice that $I^{\text{edge}}$ of two adjacent edges may differ by an integer. To
see this formally, let us consider a charge flow $\Delta Q^{\text{corner}}$ into a corner
surrounded by a closed curve $\bm x_\alpha$ with $\bm x_1=\bm x_0$ (see
Fig.~\ref{fig:Qboundary}). The net charge flow in the process is given by the
second Chern number
\begin{equation}
	\label{eq:Qc}
\Delta Q^{\text{corner}}\equiv T\oint d\bm{x}_\alpha\times\mathbcal{j}(\bm{x}_\alpha)\cdot\bm{e}_z=\int \frac{d\theta d\phi}{2\pi}G_2(\theta,\phi).
\end{equation}
For example, when the corner is formed by two edges along $x$ and $y$
directions, we have 
\begin{equation}
\Delta Q^{\text{corner}}=T(I_x^{{\text{edge}}}-I_y^{\text{edge}}),
\label{n1n2}
\end{equation}
meaning that the charge transfer along two intersecting edges can only differ
by an integer multiple of $e$. Clearly, $\Delta Q^{\text{corner}}$ is \textit{not} a bulk
topological invariant in general, since its value can be changed by closing the
boundary gap, i.e., attaching 1D Thouless pump at certain boundaries
(Figs.~\ref{fig:Qboundary} and~\ref{fig:C4}b).  

\subsection{Symmetry constraints and corner charge}
\label{sec:symmetries}
Here we consider adiabatic process of two-dimensional systems constrained by
certain symmetries that quantize $T\mathcal{m}_z^\text{top}$. We show that, if
the symmetry allows one to define the bulk contribution to quadrupole moment,
the quantized quadrupole moment is equal to $T\mathcal{m}_z^\text{top}$. Such adiabatic
processes were recently discussed by van Miert and Ortix, who found the
connection between the quantized corner charge and higher-order topological
invariant.~\cite{vanmiert2018}

\begin{figure}
	\begin{center}
		\includegraphics[width=\columnwidth]{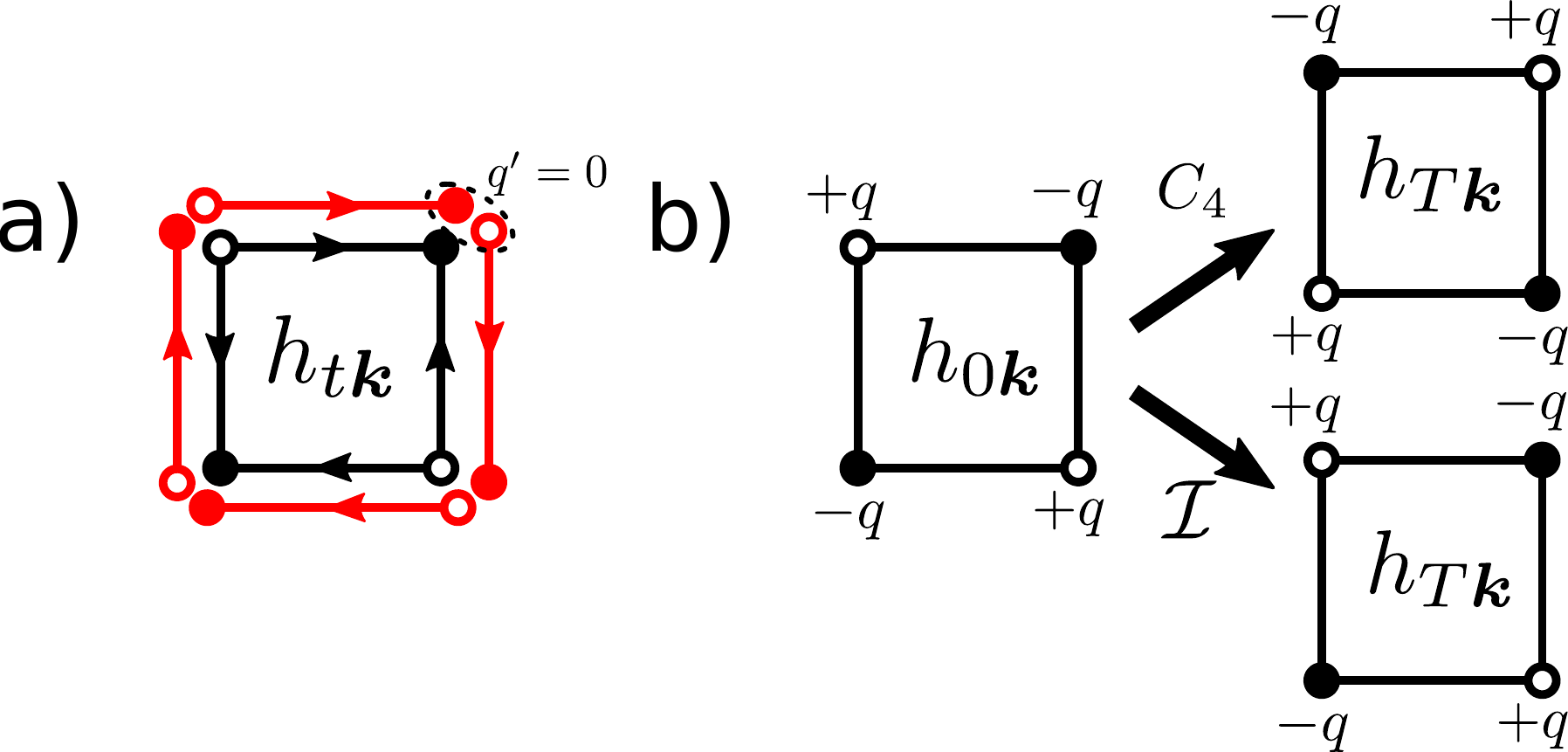}
		\caption{\label{fig:C4}  a): The four-fold rotation symmetry $C_4$ of $h_{0\bm{k}}$ imposes constraint on
		the time-independent boundary decorations and they cannot alter the corner
		charge. b): Decorating the boundary with one-dimensional
		Thouless pumps while respecting $C_4$ symmetry (combined with the time flip) of the adiabatic process can change $\Delta
		Q^\text{corner}$ by an \textit{even} integer. c): Comparison of
		action of $C_4$ and the inversion $\mathcal{I}$ on the corner charge distribution
		after one period of adiabatic process.}
	\end{center}
\end{figure}

For concreteness, let us consider the four-fold rotation $C_4$ mapping
$\bm{x}=(x,y,0)$ to $C_4\bm{x}=(-y,x,0)$. It is easy to see that boundary
decorations by polarized one-dimensional chains do not affect the fractional
part of the corner charge $\Delta Q^{\text{corner}}$, see Fig.~\ref{fig:C4}a.  We
consider an \textit{arbitrary} interpolation between the system of interest
$h_{0\bm k}$ and the reference system $h_{T/2\,\bm k}$ that has no
corner charge. The second half of the cycle is performed in a $C_4$-symmetric
manner
\begin{equation}
	U_{C_4}h_{t\bm k}U_{C_4}^\dagger=h_{T-t\,C_4\bm k}.
	\label{eq:C4cycle}
\end{equation}
The $C_4$ symmetry defined above behaves as the roto-inversion $IC_4$ in
$(t,k_x,k_y)$-space, resulting in the following transformation law for
$\mathcal{m}_z^\text{top}$:
\begin{equation}
	C_4:\quad \mathcal{m}_z^\text{top}\rightarrow-\mathcal{m}_z^\text{top}.
	\label{eq:P3_C4}
\end{equation}
This does not mean that $T\mathcal{m}_z^\text{top}$ vanishes since it is defined only ${\rm
mod}\,1$.  Thus in the presence of $C_4$ symmetry $T\mathcal{m}_z$ is quantized
either $0$ or $e/2\mod e$. When $T\mathcal{m}_z=e/2\mod e$, the circulating
edge current as in Fig.~\ref{fig:Qboundary} violates $C_4$
symmetry constraint~(\ref{eq:C4cycle})---the only allowed edge current
distribution is shown by black arrows in Fig.~\ref{fig:C4}b.  Note that the
inversion symmetry, for example, also quantizes $T\mathcal{m}_z^\text{top}$ but
the total corner charge accumulation during inversion-symmetric cycles need to
vanish since the charge distribution of quadrupole moment is invariant under
the inversion (see Fig.~\ref{fig:C4}c).

Now we show that the parity of the corner charge accumulation $\Delta
Q^{\text{corner}}$ is actually a bulk topological invariant for symmetric
adiabatic processes satisfying constraint~\eqref{eq:C4cycle}, see also
Fig.~\ref{fig:C4}b. To this end, consider two perpendicular edges along $x$
and $y$ direction, related to each other by $C_4$ symmetry. The
relations~\eqref{n1n2} and \eqref{eq:P3_C4} suggest that
$I^{\text{edge}}_y=-I^{\text{edge}}_x$ and that
\begin{equation}
	\Delta Q^{\text{corner}}=2TI^{\text{edge}}_x=2T\mathcal{m}_z^\text{top}(\bm x_0)\mod 2e.
	\label{eq:qcorner}
\end{equation}
Furthermore, Fig.~\ref{fig:C4}c tells us that the corner charge accumulation
during the symmetric process is $\Delta Q^{\text{corner}}=2 q^{\text{corner}}$. Therefore,
\begin{equation}
	q^{\text{corner}}=T\mathcal{m}_z^\text{top}(\bm x_0) =P_3(\phi_0) \mod e.\label{corner}
\end{equation}
We will discuss an example  of quadrupole insulators with $P_3=e/2$ in
Sec.~\ref{subsec:tgeoM} using this result. On the other hand, a $C_4$-symmetric
phase that hosts a corner charge of $q^{\text{ corner}}=e/4$ were recently
reported.~\cite{benalcazar2018} The fact that $\Delta Q^{\text{
corner}}\in\mathbb{Z}$ forces us to conclude that $C_4$-symmetric
adiabatic process cannot be constructed for such a phase.

Alternatively, as discussed in detail in Ref.~\onlinecite{vanmiert2018}, the
$C_4$-symmetric adiabatic process $h_{t\bm k}$ considered above, can be
viewed as a 3D topological insulator protected by the roto-inversion symmetry
$IC_4$ upon identification $k_z=2\pi t/T$. In fact, the 3D topological insulator
with $P_3=e/2$ obtained this way is a second-order topological insulator. If we
consider a geometry with the open boundary conditions in $xy$-plane and the
periodic boundary conditions in $z$-direction, such a second-order phase can be
translationally invariant in $z$-direction both in the bulk and on the boundary.
The boundary hosts an odd number of chiral modes running along each of four
hinges in $IC_4$-symmetric manner. Going back to the picture of an adiabatic
process, it becomes clear that the corner charge accumulation is an odd integer
as $t$ is varied from $0$ to $T$, which is consistent with the above
result~(\ref{eq:qcorner}).

\section{Examples: noninteracting systems}\label{sec:examples}
In this section, we discuss a simple model of noninteracting spinless electrons
in a periodic potential, which highlights the distinction of two contributions
to the bulk orbital magnetization, $\bm{m}_{\text{pers}}$ and $\mathbcal{m}$.
Additionally, we want to consider examples where there is only topological
geometric magnetization, Sec.~\ref{subsec:tgeoM}, only non-topological
geometric magnetization, Sec.~\ref{subsec:nontgeoM}, and both topological and
non-topological contributions, Sec.~\ref{subsec:rotoM}. To keep the discussions
simple while capturing the relevant physics, we focus on isolated orbitals
without any overlap between them.

\subsection{Bloch functions in the localized limit}
Let us consider a time-dependent deep potential $v_t^0(\bm{x})$ centering at
$\bm{x}=\bm{r}(t)$ that accommodates at least one bound state.  Let $\phi_t^0(\bm x)$ be the wavefunction of the instantaneous lowest-energy bound state, satisfying
$h_t^0\phi_t^0(\bm{x})=\varepsilon_t^0\phi_t^0(\bm{x})$ with 
\begin{equation}
h_{t}^0=\frac{1}{2m}[\tfrac{1}{i}\bm{\nabla}-e\bm{A}_t^{\text{ex}}(\bm{x})]^2+v_t^0(\bm{x}).\label{simplemodel}
\end{equation}
Here $\bm{A}_t^{\text{ex}}(\bm{x})$ describes an external field.  In these expressions, the superscript $0$ implies the quantities for an isolated orbit.  
When the
potential $v_t^0(\bm{x})$ is deep enough, $\phi_t^0(\bm{x})$ should be
well-localized around $\bm{x}=\bm{r}(t)$ with the localization length $\xi\ll a$.
Hence, we assume that 
\begin{equation}
\int d^2x |\phi_t^0(\bm{x})|^2=1,\quad \int d^2x \bm{x}|\phi_t^0(\bm{x})|^2=\bm{r}(t)\label{exp}
\end{equation}
and that both $\phi_t^0(\bm{x})$ and $v_t^0(\bm{x})$ decays fast enough, i.e.,
$|v_t^0(\bm{x})|,|\phi_t^0(\bm{x})|\rightarrow0$ as $|\bm{x}-\bm{r}(t)|\gg\xi$.  

With these building blocks, we construct a periodic potential and the
cell-periodic Bloch state.
\begin{align}
&v_t(\bm{x})\equiv\sum_{\bm{R}}v_t^0(\bm{x}-\bm{R}),\label{pot}\\
&u_{t\bm k}(\bm{x})\equiv\frac{a}{\sqrt{V}}\sum_{\bm{R}}e^{i\bm{k}\cdot(\bm{R}-\bm{x})}\phi_t^0(\bm{x}-\bm{R}),\label{eq:Bloch}
\end{align}
We assume that $\bm{A}^{\text{ex}}(\bm{x})$ respects the
periodicity, i.e.,
$\bm{A}^{\text{ex}}(\bm{x}-\bm{R})=\bm{A}^{\text{ex}}(\bm{x})$.  Then, as far
as $\phi_t^0(\bm{x}-\bm{R})^*\phi_t^0(\bm{x})$ and
$v_t^0(\bm{x}-\bm{R})\phi_t^0(\bm{x})$ ($\bm{R}\neq\bm{0}$) are entirely neglected,
$u_{t\bm k}(\bm{x})$ is an eigenstate of the periodic Hamiltonian
\begin{equation}
h_{t\bm{k}}=\frac{1}{2m}[\tfrac{1}{i}\bm{\nabla}-e\bm{A}_t^{\text{ex}}(\bm{x})+\bm{k}]^2+v_t(\bm{x})\label{egH}
\end{equation}
with a completely flat band dispersion $\varepsilon_{t\bm{k}}=\varepsilon_t^0$. 

\subsection{Polarization and instantaneous magnetization}
Let us first demonstrate the modern theory formula for the polarization and the
orbital magnetization by deriving $\bm{p}$ and $\bm{m}$ in two different ways.

First, we present direct calculation of the polarization and the orbital
magnetization from the microscopic charge distribution and the persistent
current densities in this insulator. The instantaneous contribution to the
local charge and current distribution from a single orbit $\phi_t^0(\bm{x})$
can be written as
\begin{align}
	\label{eq:nt0}
&n_t^0(\bm{x})\equiv e|\phi_t^0(\bm{x})|^2,\\
&\bm{j}_t^0(\bm{x})\equiv\frac{e}{m i}\phi_t^0(\bm{x})^*(\bm{\nabla}-ie\bm{A}^{\text{ex}}(\bm{x}))\phi_t^0(\bm{x}).
\label{eq:jt0}
\end{align}
We introduce vector fields $\bm{p}_t^0(\bm{x})$ and $\bm{m}_t^0(\bm{x})$ such
that
\begin{equation}
n_t^0(\bm{x})=\bar{n}^0-\bm{\nabla}\cdot\bm{p}_t^0(\bm{x}),\quad
\bm{j}_t^0(\bm{x})=\bm{\nabla}\times\bm{m}_t^0(\bm{x}).\label{defm0}
\end{equation}
The existence of such $\bm{m}_t^0(\bm{x})$ is guaranteed by the divergence-free
nature of the instantaneous current density $\bm{j}_t^0(\bm{x})$. The current
density induced by the adiabatic motion of $\bm{r}(t)$ is captured by
$\mathbcal{j}_t^0(\bm{x})$ in Eq.~\eqref{idcurrent} whose divergence may not
vanish. We assume both $\bm{p}_t^0(\bm{x})$ and $\bm{m}_t^0(\bm{x})$ decay
rapidly for $|\bm{x}-\bm{r}(t)|>\xi$, which specifies the boundary
condition for differential equations~(\ref{defm0}).

Physical quantities of the insulator composed of periodically arranged
localized orbits can be written as the sum of the contributions from each
orbit. For example microscopic current is given by
\begin{equation}
\bm{j}^{\text{micro}}(t,\bm{x})\equiv\sum_{\bm{R}}\bm{j}_t^0(\bm{x}-\bm{R})
\end{equation}
and analogously for $n$, $\bm{p}$, and $\bm{m}$. These \textit{microscopic}
expressions have a strong spatial dependence, periodically oscillating at the
scale of $a$.  To derive to a smooth mesoscopic description, we need to perform
a convolution in space (Sec. 6.6 of Ref.~\onlinecite{jackson1999}). Here we
choose the Gaussian $g(\bm{x})=(\pi R^2)^{-1}e^{-|\bm{x}|^2/R^2}$ ($R\gg a$)
\begin{equation}
\bm{j}(t,\bm{x})\equiv\int d^2x' g(\bm{x}-\bm{x}')\bm{j}^{\text{micro}}(t,\bm{x}').
\end{equation}
We do the same for other quantities. Relations such as
$\bm{j}(t,\bm{x})=\bm{\nabla}\times\bm{m}(t,\bm{x})$ are preserved by the
convolution. Because the convolution is identical to the average for the
periodic distribution, we find $n(t,\bm{x})=\bar{n}=\frac{e}{a^2}$,
$\bm{j}(t,\bm{x})=\bm{0}$,
\begin{align}
&\bm{p}(t,\bm{x})=\frac{1}{a^2}\int d^2x'\bm{p}_t^0(\bm{x}'),\label{modd2}\\
&\bm{m}_{\text{pers}}(t,\bm{x})=\frac{1}{a^2}\int d^2x'\bm{m}_t^0(\bm{x}').\label{mod2}
\end{align}
 This is the part of the orbital magnetization produced by the persistent current as illustrated in Fig.~\ref{fig:magnetization}a.

Let us check that we get the same results using the general formulae of the
modern theory. Because of the non-overlapping assumption of $\phi_t^0(\bm{x})$,
it can be readily shown that the formula in Eqs.~\eqref{eq:modthp},
\eqref{eq:modthm} for the Bloch function~\eqref{eq:Bloch} can be simplified to 
\begin{align}
&\bm{p}(t)=\frac{1}{a^2}\int d^2x\bm{x}n_t^0(\bm{x})\,\,\left(=\frac{e}{a^2}\bm{r}(t)\right),\label{modd1}\\
&\bm{m}_{\text{pers}}(t)=\frac{1}{2a^2}\int d^2x\bm{x}\times\bm{j}_t^0(\bm{x}).\label{mod1}
\end{align}
where we used Eqs.~\eqref{eq:nt0} and \eqref{eq:jt0}. These are well-known
expressions in classical electrodynamics for the charge and current
distributions in a confined region (see Sec.~4.1 and 5.6 of
Ref.~\onlinecite{jackson1999}). The equivalence of Eqs.~\eqref{modd2},
\eqref{mod2}  and \eqref{modd1}, \eqref{mod1} can be easily checked by using
the definition of $\bm{p}_t^0$ and $\bm{m}_t^0$ in Eq.~(\ref{defm0}) and
integrating by parts. The second equality of Eq.~\eqref{modd1} follows from
Eqs.~\eqref{exp}, \eqref{eq:nt0}, and \eqref{modd1}.

\subsection{Topological geometric magnetization}\label{subsec:tgeoM}
Next, we discuss the topological geometric contribution
$\mathbcal{m}^{\text{top}}$ for this model. To this end, suppose that the
position of the potential minimum $\bm{r}(t)$ adiabatically moves as a function
of  $t\in[0,T]$ and forms a closed curve as illustrated in
Fig.~\ref{fig:magnetization}b. We assume the form of the potential, and thus
the localization length, remains unchanged during the adiabatic process.

We first apply our general expression for $\mathbcal{m}^{\text{top}}$ in
Eq.~\eqref{eq:topc} to the Bloch function~\eqref{eq:Bloch}. Thanks to the
non-overlapping assumption, the vector potential $\bm{A}_{(t,\bm{k})}$ is
$\bm{k}$-independent:
\begin{equation}
	\bm{A}_{\bm{K}}=(A_t,-\bm{r}(t))^{\rm T}\label{eq:AK}
\end{equation}
with $A_t\equiv -i\int d^2x\phi_t^0(\bm{x})^*\partial_t\phi_t^0(\bm{x})$.
Plugging this into Eq.~\eqref{eq:3Dtopo2}, we find
\begin{equation}
P_3=\frac{e}{2a^2}\int_0^{T} dt\,\bm{r}(t)\times \partial_t\bm{r}(t)\cdot\bm{e}_z=\frac{eS_{\bm{r}}}{a^2},
\end{equation}
where $S_{\bm{r}}$ represents the area enclosed by the orbit of $\bm{r}(t)$ in
one cycle. Therefore,
\begin{equation}
	\mathbcal{m}^{\text{top}}=\bm{e}_z\frac{eS_{\bm{r}}}{Ta^2}=\frac{e}{2Ta^2}\oint \bm{r}(t)\times d\bm{r}(t).\label{MArea1}
\end{equation}
Observe the analogy to $\bm{m}$ in Eq.~\eqref{mod1}. This expression does not
have integer ambiguity because it is given by \textit{abelian} third
Chern-Simons form.

\begin{figure}
	\begin{center}
		\includegraphics[width=0.5\columnwidth]{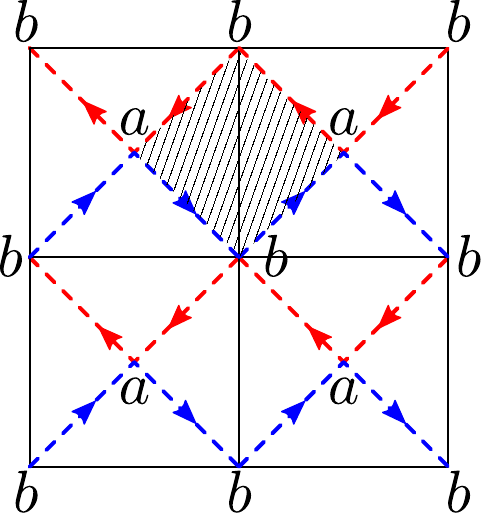}		
		\caption{\label{fig:C4e}  An adiabatic process with four-fold
		rotational symmetry $C_4$.  Each unit cell contains two
		occupied Wannier orbitals, whose trajectories during adiabatic
		process are shown with dashed red and blue lines. The hatched
		area is Arahonov-Bohm flux per unit cell acquired by such
		adiabatic process under applied magnetic field. Letters $a$
		and $b$ denote Wyckoff positions.}
	\end{center}
\end{figure}

Let us verify this result from a direct calculation. The adiabatic motion of
the single orbit following the potential minimum at $\bm{x}=\bm{r}(t)$ induces
a local current distribution
\begin{equation}
\mathbcal{j}_t^0(\bm{x})=\partial_t\bm{r}(t)n_t^0(\bm{x}).\label{idcurrent}
\end{equation}
It becomes divergence-free if averaged over one period
\begin{align}
&\mathbcal{j}^0(\bm{x})=\int_0^T\frac{dt}{T}\mathbcal{j}_t^0(\bm{x})=\frac{1}{T}\oint  d\bm{r}(t)n_t^0(\bm{x}),\\
&\bm{\nabla}\cdot\mathbcal{j}^0(\bm{x})=-\frac{e}{T}\oint  d\bm{r}\cdot\bm{\nabla}_{\bm{r}}n_t^0(\bm{x})=0.
\end{align}
As we have seen above, the sum of such microscopic currents from each unit cell
produces the bulk magnetization
\begin{align}
\mathbcal{m}^{\text{top}}&=\frac{1}{2a^2}\int d^2x\,\bm{x}\times\mathbcal{j}^0(\bm{x})\notag\\
&=\frac{e}{2Ta^2}\oint \left(\int d^2x \bm{x}|\phi_0(\bm{x}-\bm{r}(t))|^2\right)\times d\bm{r}(t).
\end{align}
This agrees with Eq.~\eqref{MArea1} because the integral in the parenthesis is
precisely $\bm{r}(t)$ due to Eq.~\eqref{exp}.

The result in Eq.~\eqref{MArea1} can be readily generalized to the case with
multi-orbitals, such as examples in Fig.~\ref{fig:magnetization}c and e. Let
us introduce potential minima $\bm{x}=\bm{r}_n(t)$ ($n=1,2,\cdots$) in each
unit cell, which are adiabatically varied as a function of $t\in[0,T]$. This
time, each orbit is allowed to form an \textit{open} curve, as far as the total
polarization $\bm{p}(t)=(e/a^2)\sum_{n}\bm{r}_n(t)$ satisfies
$\bm{p}(T)=\bm{p}(0)$.  Under such an assumption, we find that
\begin{align}
	\label{eq:P3ukN}
	\mathbcal{m}^{\text{top}}=&\sum_{n}\frac{e}{Ta^2}\bigg(\bm{S}_{\bm{r}_n}+\frac{1}{2}\bm{r}_n(0)\times\bm{r}_n(T)\bigg),\\
	\bm S_{\bm{r}_n}&\equiv\frac{1}{2}\int_{\bm{r}_n(0)}^{\bm{r}_n(T)}\bm{r}_n(t)\times d\bm{r}_n(t).\label{Sn}
\end{align}
We present the proof in the Appendix~\ref{app:MzKP}.  Although the above
expression appears to be the sum of single-band contributions, the ``would-be''
contribution from each band depends on the specific choice of the origin when
it does not form a closed loop.  Only after performing the summation over
all occupied bands, or in other words, only after fully taking into account the
\textit{non-abelian} nature of the third Chern-Simons form, the result restores
the independence from the origin choice.

As the application of the formula \eqref{eq:P3ukN}, let us discuss the corner
charge of the $C_4$-symmetric quadrupole insulator introduced in
Refs.~\onlinecite{benalcazar2017,benalcazar2017}.  
For the wallpaper group
$p4$, there exist three spacial Wyckoff positions: the unit cell origin at
$\bm{x}_a=(0,0)$, the center of the plaquette at $\bm{x}_b=(a/2,a/2)$, and the
center of bonds at $\bm{x}_c=(a/2,0)$, $(0,a/2)$.~\cite{ITC}  In the nontrivial
phase, the two occupied Wannier orbitals locate at $\bm{x}_b$, while in the
trivial phase they are at $\bm{x}_a$. We consider a periodic adiabatic process
illustrated in Fig.~\ref{fig:C4e} starting with the nontrivial phase at $t=0$
and passing the trivial phase at $t=T/2$. The
instantaneous Hamiltonian $h_{t\bm{k}}$ itself breaks the $C_4$-symmetry down
to $C_2$ symmetry except when $t=0$ and $T/2$, while the adiabatic process as a
whole implements the full $C_4$ in the sense of Eq.~\eqref{eq:C4cycle}.  We can
readily compute $P_3$ of this process using Eq.~\eqref{eq:P3ukN} which turns
out to be $e/2$. This is the corner charge of the quadrupole insulator as
predicted by Eq.~\eqref{corner}, which agrees with the original
study.~\cite{benalcazar2017,benalcazar2017} A variant of this adiabatic
process was also discussed in Ref.~\onlinecite{vanmiert2018}.

\subsection{Non-topological geometric magnetization}\label{subsec:nontgeoM}
In this example we first consider a single electron in an anisotropic and
rotating two-dimensional well,~\cite{goldman2014} see
Fig.~\ref{fig:magnetization}d. We assume a harmonic confining potential, i.e.,
Hamiltonian~(\ref{simplemodel}) with 
\begin{equation}
	v_t^0(\bm x)=\frac{1}{2}m(\omega_x^2x_t^2+\omega_y^2y_t^2),
	\label{eq:assymv0}
\end{equation}
where $\bm x_t\equiv(x\cos\Omega t+y\sin\Omega t,-x\sin\Omega t+y\cos\Omega t)$
and $\Omega=2\pi/T$. To obtain the geometric orbital magnetization for this
model, we consider external magnetic field $\bm B=B_z\bm e_z$ described by the
vector potential $\bm A^{\rm ex}(\bm x)=\bm B\times\bm x/2$. (Strictly speaking,
this form is valid only around the origin as it lacks the required
periodicity.) The wave function of the instantaneous ground state of this model
can be obtained based on Ref.~\onlinecite{rebane2012}:
\begin{align}
	\phi_{t}^0(\bm x)=\mathcal{N}e^{\frac{im\omega_c(\omega_y-\omega_x)x_ty_t}{2(\omega_x+\omega_y)}-\frac{m\sqrt{(\omega_x+\omega_y)^2+\omega_c^2}(\omega_xx_t^2+\omega_yy_t^2)}{2(\omega_x+\omega_y)}},
	\label{eq:phi0assym}
\end{align}
where $\cal N$ is the normalization factor and $\omega_c\equiv eB_z/m$ is the
cyclotron frequency. The Berry phase $\varphi_{B_z}$ during the adiabatic
process $t\in[0,T]$ is
\begin{align}
	\varphi_{B_z}^0&=\int_0^{T}dt\int d^2 x\phi_{t}^0(\bm x)^*i\partial_t\phi_{t}^0(\bm x)\notag\\
	&=\frac{\pi \omega_c(\omega_y-\omega_x)^2}{2\omega_x\omega_y\sqrt{(\omega_x+\omega_y)^2+\omega_c^2}}.
	\label{eq:phiberry}
\end{align}
From Eq.~(\ref{eq:calMz}) it follows that the adiabatic
process~(\ref{eq:assymv0}) has non-zero geometric orbital magnetic moment
$\mathcal{m}_z^0$
\begin{align}
		\mathcal{m}_z^0=\frac{e(\omega_y-\omega_x)^2\Omega}{4ma^2\omega_x\omega_y(\omega_x+\omega_y)}.\label{cmz0}
\end{align}

Now we construct the Bloch function \eqref{eq:Bloch} using $\phi_{t}^0(\bm x)$
as the building block and compute the geometric orbital magnetization
$\mathcal{m}_z$ based on Eq.~\eqref{eq:nontopcalM} for the corresponding band
insulator.  To this end we need  instantaneous eigenstates and eigenenergies in
absence of external magnetic field including unoccupied bands.  The Hamiltonian
$h_{t\bm k}$ is given by Eq.~(\ref{egH}) with $v_t^0(\bm x)$ given by
Eq.~(\ref{eq:assymv0}) and $\bm A^{\rm ex}=0$.  We assume that there is no
overlap between wavefunctions belonging to different unit cells as before.
(When $\omega_{x}, \omega_{y}$ are large enough, such an assumption is valid at
least for relevant low-energy states.) Bloch
wavefunctions read 
\begin{align}
u_{t\bm{n}\bm k}(\bm{x})&\equiv\frac{a}{\sqrt{V}}\sum_{\bm{R}}e^{i\bm{k}\cdot(\bm{R}-\bm{x})}\phi_{t\bm{n}}^0(\bm{x}-\bm{R}),
\end{align}
where $\bm{n}\equiv(n_x,n_y)$ labels energy levels of two-dimensional the
anisotropic harmonic oscillator and $\bm{n}=(0,0)$ corresponds to the ground
state in Eq.~\eqref{eq:phi0assym} with $\omega_c=0$. Substituting above
expressions to Eqs.~(\ref{eq:topc}) and~(\ref{eq:nontopcalM}), we find
\begin{align}
	&\mathcal{m}_z^{\text{top}}=0,\\
	&\mathcal{m}_z^{\text{non-top}}=\sum_{\bm{n}\neq(0,0)}\frac{\vert\langle\phi_{t}^0\vert\bm x\times\bm\nabla\vert\phi_{t\bm{n}}^0\rangle\vert^2e\Omega}{4ma^2(n_x\omega_x+n_y\omega_y)}\notag\\
	&=\frac{\vert\langle\phi_{t}^0\vert\bm x\times\bm\nabla\vert\phi_{t(1,1)}^0\rangle\vert^2e\Omega}{4ma^2(\omega_x+\omega_y)}=\mathcal{m}_z^0.
	\label{eq:barnett}
\end{align}

\subsection{Geometric magnetization by rotation}\label{subsec:rotoM}
Here we calculate the contribution to the geometric orbital magnetization of a rotating
uncharged body and compare it to the classical Barnett
effect.~\cite{barnett1915,barnett1935} The Barnett effect predicts magnetization
$\chi/\gamma\Omega$, where $\chi$ is the paramagnetic susceptibility, $\gamma$
is the electron gyromagnetic ratio, and $\Omega$ is the rotation frequency. Since the rotation
axis does not necessarily coincide with potential well minima we have $v_t^0(\bm x-\bm
r(t))$ with $v_t^0$ from Eq.~(\ref{eq:assymv0}) and $\bm r(t)=(R\cos\Omega
t,R\sin\Omega t,0)^{\rm T}$, where $R$ is the distance of the potential well minima to the
rotation axis. The lowest-energy instantaneous wavefunction $\phi_t(\bm x)$
can be obtained from Eq.~(\ref{eq:phi0assym}) by performing gauge
transformation
\begin{align}
	\label{eq:phi0}
	\phi_{t}^0(\bm x)=\mathcal{N}e^{\frac{im\omega_c(\omega_y-\omega_x)(x_t-R)y_t}{2(\omega_x+\omega_y)}+\frac{i}{2}m\omega_c\bm{e}_z\cdot\bm r(t)\times\bm x}\\
	\times e^{-\frac{m\sqrt{(\omega_x+\omega_y)^2+\omega_c^2}(\omega_x(x_t-R)^2+\omega_yy_t^2)}{2(\omega_x+\omega_y)}}.\notag
\end{align}
As compared to Eq.~\eqref{eq:phiberry}, the Berry phase $\varphi_{B_z}$ during
the adiabatic process $t\in[0,T]$ acquires an additional contribution $eB_z \pi
R^2$ from the  Aharonov-Bohm phase. Therefore electrons contribute to the
following geometric orbital magnetization 
\begin{align}
	\mathcal{m}_z&=\frac{eR^2\Omega}{2a^2}+\mathcal{m}_z^0.
\end{align}
The first term can be identified with $\mathcal{m}_z^{\text{top}}$ in
Eq.~(\ref{MArea1}) and the second term is the contribution in Eq.~\eqref{cmz0}.
Since the body is uncharged, the contribution from ions cancels the topological
contribution, while $\mathcal{m}_z^{\text{non-top}}=\mathcal{m}_z^0$ remains
since ions are much more localized compared to electrons. Assuming anisotropy
$\omega_x/\omega_y=2$, and confinement of electrons on the scale of angstroms,
we find that contribution~(\ref{eq:barnett}) is on the same order as Barnett
effect for paramagnets with paramagnetic susceptibility $\chi\sim10^{-5}$. For
comparison, paramagnets have typically magnetic susceptibility
$\chi\sim10^{-3}-10^{-5}$.~\cite{wiki:paramagnetism}

\section{Examples: finite interacting systems}\label{sec:examples2}
In this section we demonstrate the validity of Eq.~\eqref{eq:calMz} for
finite interacting systems. We consider two canonical ways of introducing the
time-dependence to the Hamiltonian: rotating~\cite{ceresoli2002,stengel2018}
and translating the whole system.~\cite{goldman2014,juraschek2017,dong2018}

We consider many-body systems under the \textit{open boundary condition} in two
spatial dimensions.  We start with a \textit{time-independent} Hamiltonian
$\hat{H}$ that can contain arbitrary interactions.  The total charge, current,
polarization, and orbital magnetization operator for this Hamiltonian can be
written as $\hat{\bm{N}}=\int_Vd^2x \hat{\bm{n}}(\bm{x})$, $\hat{\bm{J}}=\int
d^2x \hat{\bm{j}}(\bm{x})$, $\hat{\bm{X}}=\int_Vd^2x
\bm{x}\hat{\bm{n}}(\bm{x})$, $\hat{\bm{M}}=(1/2)\int_Vd^2x
\bm{x}\times\hat{\bm{j}}(\bm{x})$.  We stress that these expressions are valid
only when the system is confined in a finite region; they need to be modified
in extended systems under the periodic boundary conditions as done by the
modern theory.  We denote the many-body ground state of $\hat{H}$ and its
energy by $|\Phi\rangle$ and $E$, respectively.

To compute the many-body Berry phase, let $\hat{H}_{\bm{B}}$ be the Hamiltonian
with the vector potential in the symmetric gauge
$\bm{A}^{\text{ex}}(\bm{x})=(1/2)\bm{B}\times\bm{x}$ with $\bm{B}=B_z\bm{e}_z$.
Expanding to the linear order in $B_z$ and using
$\hat{\bm{j}}(\bm{x})=-\partial_{\bm{A}(\bm{x})}\hat{H}$, we get
\begin{equation}
\hat{H}_{B_z}=\hat{H}-\hat{M}_zB_z+O(B_z^2).\label{perturbationmB}
\end{equation}
Therefore, the ground state of $\hat{H}_{B_z}$ to the leading order in $B_z$
can be expressed as
\begin{equation}
|\Phi_{B_z}\rangle=|\Phi\rangle+\hat{Q}\frac{1}{\hat{H}-E}\hat{Q}\hat{M}_zB_z|\Phi\rangle+O(B_z^2).\label{firstB}
\end{equation}
Here $\hat{Q}\equiv1-|\Phi\rangle\langle\Phi|$ is the projector onto excited
states.

\subsection{Rotation}\label{subsec:rotom}
Here we consider the time-dependence of the problem induced by the rotation of
the whole system
\begin{equation}
\hat{H}_t\equiv e^{-i\hat{L}_z\Omega t}\hat{H}e^{i\hat{L}_z\Omega  t},
\end{equation}
where $\Omega=\bm{e}_z\Omega$ is the rotation frequency and $\hat{\bm{L}}$ is
the angular momentum operator. For the time-dependent Hamiltonian $\hat{H}_t$, 
the orbital magnetization operator $\hat{\bm{M}}_t\equiv (1/2)\int_Vd^2x
\bm{x}\times\hat{\bm{j}}_t(\bm{x})$
\begin{equation}
\hat{\bm{M}}_t=e^{-i\hat{L}_z\Omega t}\hat{\bm{M}}e^{i\hat{L}_z\Omega t}.\label{eq:rotm}
\end{equation}

We evaluate the instantaneous contribution $\bm{m}_{\text{pers}}$ and the
geometric contribution $\mathbcal{m}$ to the orbital magnetization via the
formulae in Eqs.~\eqref{eq:Xinst} and \eqref{eq:Xgeom}. The instantaneous
contribution is given by the instantaneous ground state
$|\Phi_t\rangle=e^{-i\hat{L}_z\Omega  t}|\Phi\rangle$
\begin{equation}
\bm{m}_{\text{pers}}\equiv\int_0^T\frac{dt}{T}\frac{\langle\Phi_t|\hat{\bm{M}}_t|\Phi_t\rangle}{V}=\frac{\langle\Phi|\hat{\bm{M}}|\Phi\rangle}{V}.\label{M1}
\end{equation}
The geometric contribution is given by the many-body Berry phase.  Since the
instantaneous ground state of the Hamiltonian $\hat{H}_{tB_z}\equiv
e^{-i\hat{L}_z\Omega  t}\hat{H}_{B_z}e^{i\hat{L}_z\Omega  t}$ is given by
$|\Phi_{tB_z}\rangle=e^{-i\hat{L}_z\Omega  t}|\Phi_{B_z}\rangle$, we have 
\begin{equation}
\varphi_{B_z}=\int_0^Tdt\langle\Phi_{tB_z}|i\partial_t|\Phi_{tB_z}\rangle=T\langle\Phi_{B_z}|\hat{L}_z\Omega|\Phi_{B_z}\rangle.
\end{equation}
This is the expectation value of $\hat{L}_z\Omega$ in the presence of the
perturbation $-\hat{m}_z B_z$ in Eq.~\eqref{perturbationmB}. Using
Eq.~\eqref{firstB}, we get
\begin{equation}
	\mathbcal{m}=\langle\Phi|\hat{L}_z\Omega\hat{Q}\frac{1}{\hat{H}-E}\hat{Q}\frac{\hat{\bm{M}}}{V}|\Phi\rangle+\text{c.c.}\label{M2}
\end{equation}

We verify these results by solving time-dependent problem.  The solution to the
\textit{time-dependent} Schr\"odinger equation
$i\partial_t|\Psi_t\rangle=\hat{H}_t|\Psi_t\rangle$ can be readily constructed
using the ground state $|\Phi_{\Omega}\rangle$ of the \textit{time-independent}
Hamiltonian
\begin{equation}
\hat{H}_{\Omega}\equiv \hat{H}-\hat{L}_z\Omega.\label{perturbationOL}
\end{equation}
The solution that is smoothly connected to the ground state in the static limit $\Omega\rightarrow0$ reads
\begin{equation}
|\Psi_t\rangle=e^{-i\hat{L}_z\Omega  t-iE_{\Omega}t}|\Phi_{\Omega}\rangle.\label{eq:rotphi2}
\end{equation}
The time-average of the orbital magnetization is thus given by
\begin{equation}
\bm{m}=\int_0^T\frac{dt}{T}\frac{\langle\Psi_t|\hat{\bm{M}}_t|\Psi_t\rangle}{V}=\frac{\langle\Phi_{\Omega}|\hat{\bm{M}}|\Phi_{\Omega}\rangle}{V}.
\end{equation}
This is the expectation value of $\hat{\bm{M}}$ in the presence of the
perturbation $-\hat{L}_z\Omega$ as in Eq.~\eqref{perturbationOL}.  The
first-order perturbation theory with respect to $\Omega$ gives
\begin{equation}
	\bm{m}=\bm m_{\text{pers}}+\langle\Phi|\hat{L}_z\Omega\hat{Q}\frac{1}{\hat{H}-E}\hat{Q}\frac{\hat{\bm{M}}}{V}|\Phi\rangle+\text{c.c.}
\end{equation}
This is precisely $\bm{m}_{\text{pers}}+\mathbcal{m}$ predicted above in
Eqs.~\eqref{M1} and ~\eqref{M2}. As it is clear from the derivation, the
agreement of the two independent approaches is guaranteed by the Maxwell
relation for the free energy
$\hat{F}\equiv\hat{H}-\hat{L}_z\Omega-\hat{M}_zB_z$

\begin{equation}
\partial_{B_z}\langle\hat{L}_z\rangle=-\partial_{B_z}\partial_{\Omega}\langle\hat{F}\rangle=-\partial_{\Omega}\partial_{B_z}\langle\hat{F}\rangle=\partial_{\Omega}\langle\hat{M}_z\rangle.
\end{equation}

\subsection{Translation}
Next let us introduce the time-dependence by the translation.  All discussions
proceed in essentially the same way, while there are still a few differences.
First we define the time-dependent Hamiltonian by
\begin{equation}
\hat{H}_t'\equiv \hat T_t\hat{H}\hat T_t^\dagger,
\end{equation}
where $\hat T_t=e^{-i\hat{\bm P}\cdot\bm r(t)}$ is the translation by amount $\bm
r(t)$ and $\hat{\bm{P}}$ is the momentum operator. For $\hat{H}_t$  the orbital
magnetization operator becomes
\begin{equation}
\hat{\bm{M}}_t'= \hat T_t\left(\hat{\bm{M}}+\frac{1}{2}\bm{r}(t)\times\hat{\bm{J}}\right)\hat T_t^\dagger\label{eq:transm}
\end{equation}
where the second term in the parenthesis is due to the change of the origin. The
instantaneous ground state  $|\Phi_t'\rangle=\hat T_t|\Phi\rangle$
gives $\bm{m}_{\text{pers}}$ as in Eq.~(\ref{M1}), where we used
$\langle\Phi|\hat{\bm{J}}|\Phi\rangle=\bm{0}$.

Next, we compute the geometric contribution via the many-body Berry phase.
In the presence of magnetic field translation operator $\hat
T_{tB_z}\equiv\hat T_{B_z}(\bm r(t))$ becomes translation followed by gauge
transformation~\cite{brown1964,zak1964}
\begin{equation}
\partial_t\hat{T}_{tB_z}\equiv -i(\hat{\bm{P}}+\tfrac{e}{2}\bm{B}\times\hat{\bm{X}})\cdot\partial_t\bm{r}(t)\hat{T}_{tB_z}.
\end{equation}
The instantaneous ground state of $\hat{H}_{tB_z}\equiv
\hat{T}_{tB_z}\hat{H}_{B_z}\hat{T}_{tB_z}^\dagger$ is
$|\Phi_{tB_z}\rangle=\hat{T}_{tB_z}|\Phi_{B_z}\rangle$, thus the many-body
Berry phase reads
\begin{equation}
\varphi_{B_z}=\int_0^Tdt\langle\Phi_{B_z}|\hat{T}_{tB_z}^\dagger i\partial_t\hat{T}_{tB_z}|\Phi_{B_z}\rangle=eN\bm{S}_{\bm{r}}\cdot \bm{B},
\end{equation}
Here, $\bm{S}_{\bm{r}}\equiv\frac{1}{2}\oint \hat{\bm{r}}\times d\bm{r}$
represents the area swept by $\bm{r}(t)$ in one cycle. In the derivation, we
used
\begin{align}
&\hat{T}_{tB_z}^\dagger i\partial_t\hat{T}_{tB_z}\notag\\
&=(\hat{\bm{P}}+\tfrac{e}{2}\bm{B}\times\hat{\bm{X}})\cdot\dot{\bm{r}}(t)+\tfrac{eN}{2}\bm{r}(t)\times\partial_t\bm{r}(t)\cdot \bm{B}.
\end{align}
Therefore, when the whole system is translated, the geometric contribution to
the orbital magnetization captures the Aharonov-Bohm phase
\begin{equation}
\mathbcal{m}=\frac{eN}{TV}\bm{S}_{\bm{r}}.\label{M4}
\end{equation}

To verify the above results, we consider the time-dependent Schr\"odinger
equation $i\partial_t|\Psi_t'\rangle=\hat{H}_t'|\Psi_t'\rangle$. An
 approximate solution is given by $|\Psi_t'\rangle=\hat
T_t|\Phi_{t\bm{r}}\rangle$, where $|\Phi_{t\bm{r}}\rangle$ is the instantaneous
ground state of the Hamiltonian $\hat{H}_{t\bm{r}}\equiv
\hat{H}-\hat{\bm{P}}\cdot\partial_t\bm{r}(t)$. Therefore, the time-average of
the orbital magnetization is
\begin{align}
&\bm{m}=\int_0^T\frac{dt}{T}\frac{\langle\Psi_t'|\hat{\bm{M}}_t'|\Psi_t'\rangle}{V}\\
&=\int_0^T\frac{dt}{T}\frac{\langle\Phi_{t\bm{r}}|\hat{\bm{M}}|\Phi_{t\bm{r}}\rangle}{V}+\frac{eN}{2TV}\int_0^Tdt\,\bm{r}(t)\times \partial_t\bm{r}(t).\notag
\end{align}
In the adiabatic limit, this reproduces $\bm{m}_{\text{pers}}+\mathbcal{m}$ in
Eqs.~\eqref{M1} and \eqref{M4}. In the derivation we used
$\langle\Phi_{t\bm{r}}|\hat{\bm{J}}|\Phi_{t\bm{r}}\rangle=eN\partial_t\bm{r}(t)$
for the ground state of $\hat H_{t\bm r}$.

\section{Conclusion}\label{sec:conclusions}
In order to obtain current and charge distribution in a medium, one needs to
solve Maxwell's equation together with two constitutive relations [see
Eq.~(\ref{eq:jmeso})] that fully characterize the medium at the mesoscopic
scale. The modern theories, developed in the last 30 years, provide handy
formulae to calculate electric
polarization~\cite{king-smith1993,vanderbilt1993,resta1994,resta2007} and
orbital magnetization~\cite{xiao2005,thonhauser2005,ceresoli2006,shi2007} for
realistic materials.

The focus of this work is on spinless short-range entangled systems under
periodic adiabatic evolution. Our main result is to identify an additional
contribution to the orbital magnetization that we name geometric orbital
magnetization $\mathbcal{m}$. This new contribution is defined only after
performing the time-average over the period of the adiabatic process, which
makes the current density divergence-free. We find that the geometric orbital
magnetization can be expressed compactly as derivative of the many-body Berry
phase with respect to an externally applied magnetic field. For band
insulators, we obtain handy formulae for the bulk geometric orbital
magnetization $\mathbcal{m}$ in terms of instantaneous Bloch states and
energies. Interestingly, we find
that for band insulators
$\mathbcal{m}=\mathbcal{m}^{\text{top}}+\mathbcal{m}^{\text{non-top}}$ consists
of two pieces, where topological piece $\mathbcal{m}^{\text{top}}$  depends
only on the Bloch states of occupied bands. For spinless systems only electric
polarization and orbital magnetization enter constitutive relations, since the
contributions from higher moments are typically negligible.~\cite{jackson1999}
In this sense, our results together with ``the modern theories'' provide a
complete mesoscopic description of a medium under periodic adiabatic time
evolution. In this work we have not considered adiabatic processes with ground
state degeneracy,~\cite{meidan2011} it would be interesting to see to which
extent our findings can be generalized to such systems.

In the present work, the adiabaticity assumption is crucial for validity of the
obtained results. In practice, for band insulators with band gaps on the order
of electronvolt, this conditions requires that the period $T$ is larger than
couple of femtoseconds. Nevertheless, shorter period $T$ results in a larger
geometric orbital magnetization. It would be therefore interesting to extend
our results to the case of strong drive that excites unoccupied bands. In the
case of Thouless pumps such extension was very fruitful and resulted in recent
discovery of shift currents.~\cite{balz1981,morimoto2016}

Although higher (than dipole) electric and magnetic multiple moments typically
do not enter constitutive relations, the knowledge of these quantities may be
useful for certain systems.~\cite{benalcazar2017,benalcazar2018} In fact, it is
a topic of current research whether higher moments can be established as bulk
quantities in general.~\cite{gao2018b,shitade2018,kang2018,metthew2018,ono2019}
In the presence of certain crystalline symmetries, both electric
polarization and topological geometric orbital magnetization can be quantized,
in which case they can serve as a topological invariants. In this context, we
showed that the quantized quadrupole moment is related to
$\mathcal{m}^{\text{top}}_z$ in systems with proper symmetries that allow bulk
definition of the quadrupole moment.~\cite{ono2019}

In this work we succeeded in separating $\mathbcal{m}$ into the topological and
the non-topological piece only for band insulators. There, we found that the
topological contribution is expressed as the third Chern-Simons form ($P_3$), in
$(t,k_x,k_y)$ space. For interacting systems, based on examples considered in
Sec.~\ref{sec:examples2}, we conclude that it is possible to separate
Aharonov-Bohm contribution originating from the center of mass motion. In fact,
this contribution can be captured by calculating $P_3$ formally defined for the
many-body ground state as a function of time and two solenoidal fluxes.
Clearly, the many-body $P_3$ defined in such manner is abelian and does not
capture all possible topological contributions. For example, it vanishes for
the model in Fig.~\ref{fig:C4}. As a future direction, it would be interesting
to see if separation achieved for band insulators is possible for general
single-particle, or even many-body systems. The affirmative answer to this
question would provide a way to define $P_3$ in two dimensional systems with
adiabatic time-dependence lacking the translational invariance or the
single-particle description. The formula for $P_3$ in many-body
three-dimensional systems already exist in the
literature,~\cite{wang2014,shiozaki2018b} where it was argued that $P_3$ is
related to the magnetoelectric polarizability. The magnetoelectric
polarizability of three-dimensional materials contains, at
least for the case of band insulators, not only topological but also
non-topological contribution,~\cite{essin2010} thus the analogous ``separation
question'' arises also in that context. Additionally, defining quantized
quadrupole moment for interacting systems is one of the open
questions.~\cite{kang2018,metthew2018,ono2019} Since for band insulators we
find connection between $\mathcal{m}_z^{\text{top}}$ and quantized quadrupole
moment, separating $\mathcal{m}_z^{\text{top}}$ contribution in interacting
systems might provide useful many-body definition of quantized quadrupole
moment.

We hope that our work will also have practical implication as it contributes to
emerging field of ``dynamical material design'' by providing a way to calculate
additional orbital magnetization contribution that appears in these
systems.~\cite{ceresoli2002,juraschek2017,juraschek2018,dong2018,stengel2018}

\begin{acknowledgements}
We would like to thank David Vanderbilt for drawing
Refs.~\onlinecite{ceresoli2002,juraschek2017,juraschek2018,dong2018,stengel2018}
to our attention.  The work of S.O. is supported by Materials Education program
for the future leaders in Research, Industry, and Technology (MERIT).  The work
of H.W. is supported by JSPS KAKENHI Grant No.~JP17K17678 and by JST PRESTO
Grant No.~JPMJPR18LA. 
\end{acknowledgements}

\appendix
\section{Topological geometric orbital magnetization in simple models}\label{app:MzKP}
Here we present the derivation of Eq.~\eqref{eq:P3ukN}  in Sec.~\ref{subsec:tgeoM} for multi potential
minima at $\bm{x}=\bm{r}_n(t)$ ($n=1,2,\cdots,N_{\rm occ}$). To calculate the geometric orbital
magnetization, we use the
cell-periodic Bloch state~(\ref{eq:Bloch}) for each orbital.
Substitution of $(\bm A_{\bm K})_{nm}=(A_{tn},-\bm{r}_n(t))^{\rm T}\delta_{nm}$ with $\bm{K}\equiv(t,\bm{k})$ [c.f. Eq.~(\ref{eq:AK})] into Eq.~(\ref{eq:3Dtopo2}), we find
\begin{equation}
	P_3'=e\sum_{n=1}^{N_{\rm occ}}\bm{S}_{\bm{r}_n}\cdot\bm{e}_z.	
	\label{eq:P3non}
\end{equation}
However, this is not the complete expression of $P_3$ because Eq.~\eqref{eq:AK} implies that the above Berry connection violates the
periodicity in time when $\bm{r}_n(T)$ differs from $\bm{r}_n(0)$. To quantify the violation,
let us introduce the mismatch matrix
\begin{equation}
	(\Theta_{0\bm k})_{nm}\equiv\langle u_{0n\bm{k}}|u_{Tm\bm{k}}\rangle=\delta_{nm}e^{-i\bm{k}\cdot[\bm{r}_n(T)-\bm{r}_n(0)]}.
	\label{eq:M}
\end{equation}
To
repair periodicity of $\bm A_{\bm K}$ we need to find a family of
unitary matrices $\Theta_{\bm K}$ that interpolates $\Theta_{0\bm k}$ and  $\Theta_{T\bm
k}\equiv\mathds{1}_{N_{\rm occ}\times N_{\rm occ}}$. 
Such an interpolation exists if and only if
the polarization $\bm{p}(t)=(e/a^2)\sum_{n}\bm{r}_n(t)$ satisfy
$\bm{p}(T)=\bm{p}(0)$. Accordingly, below we focus on the case $\bm Q=0$.  The gauge transformation with
this $\Theta_{\bm K}$ gives a periodic Berry connection 
\begin{equation}
\tilde{\bm A}_{\bm K}=\Theta_{\bm K}^\dagger\bm A_{\bm K}\Theta_{\bm K}-i\Theta_{\bm K}^\dagger\bm{\nabla}_{\bm K}\Theta_{\bm K}
\end{equation}
as far as
\begin{equation}
\Theta_{\bm K}^\dagger\partial_{t}\Theta_{\bm K}|_{t=T}=\Theta_{\bm K}^\dagger\partial_{t}\Theta_{\bm K}|_{t=0}.
	\label{eq:Ath_P}
\end{equation}

The full expression of $P_3$ is given by computing Eq.~(\ref{eq:3Dtopo2}) with $\tilde{\bm A}_{\bm K}$. It is given by~\cite{ryu2007}
\begin{align}
&P_3=P_3'+\Delta_{\text{boundary}}+\Delta_{\text{PI}},\\
&\Delta_{\text{boundary}}=e\int \frac{d^2k}{8\pi^2}\tr [i\epsilon_{ij}\partial_{k_i}\Theta_{\bm K}\Theta_{\bm K}^\dagger(\bm A_{\bm{K}})_j]_{t=0}^{t=T},\\
&\Delta_{\text{PI}}=e\int \frac{d^3K}{24\pi^2}{\rm tr}\Theta_{\bm K}^\dagger \partial_{K_\mu}\Theta_{\bm K}\Theta_{\bm K}^\dagger\partial_{K_\nu}\Theta_{\bm K}\Theta_{\bm K}^\dagger \partial_{K_\lambda}\Theta_{\bm K},
\end{align}
where the sum over $i,j=x,y$ and $\mu,\nu,\lambda=t,x,y$ are assumed.  $\Delta_{\text{PI}}$ is the Pontryagin index of the matrix $\Theta_{\bm{K}}$, which requires an explicit expression for $\Theta_{\bm{K}}$, whereas the
knowledge of mismatch matrix $\Theta_{0\bm{k}}$ in Eq.~(\ref{eq:M}) suffices for the boundary term $\Delta_{\text{boundary}}$. It reads
\begin{equation}
\Delta_{\text{boundary}}=e\sum_{n=1}^{N_{\rm occ}}\frac{1}{2}\bm r_n(0)\times\bm r_n(T)\cdot\bm{e}_z.	
	\label{eq:P3boundary}
\end{equation}
Thus, in order to prove~(\ref{eq:P3ukN}), it remains to show that $\Delta_{\text{PI}}=0$. Below we show this by
explicitly constructing $\Theta_{\bm{K}}$.

Consider a family of $2\times2$ unitary matrices $\Xi_{\bm{K}}(\bm r)$
\begin{align}
	\Xi_{\bm{K}}(\bm r)&=e^{\frac{i}{2}\bm k\cdot\bm r\sigma_3}e^{\frac{i}{2}\bm k\cdot\bm r[\sigma_3\cos(\pi f(t))+\sigma_1\sin(\pi f(t))]}
\end{align}
with $f(0)=0$ and $f(T)=T$. The above matrix interpolates between unitary
matrix $\Xi_{0\bm{k}}(\bm r)=e^{i\sigma_3 \bm k\cdot\bm r}$ and the identity matrix $\Xi_{T\bm{k}}(\bm r)=\mathds{1}_{2\times 2}$. We also require $f^\prime(0)=f^\prime(T)=0$ so that the requirement~(\ref{eq:Ath_P}) is satisfied. One such function $f(t)$ is
given by
\begin{align}
	f(t)&=3 (t/T)^2 -2(t/T)^3.
\end{align}
Since $\Xi_{\bm{K}}(\bm r)$ depends only on $\bm k\cdot\bm r$, it is easy
to check that its Pontryagin index  vanishes. We then define $\Theta_{\bm{K}}$
by a composition 
\begin{align}
	\Theta_{\bm{K}}&=\Theta^{(0)}_{\bm{K}}\circ\Theta^{(1)}_{\bm{K}}\circ\dots\circ\Theta^{(N_{\rm occ}-1)}_{\bm{K}}, 
\end{align}
where each interpolation matrix $\Theta^{(n)}_{\bm{K}}$ is defined as,
\begin{align}
	{\rm P}_{n}\Theta^{(n)}_{\bm{K}}{\rm P}_{n}&={\rm P}_{n}\Theta^{(n-1)}_{T\bm k}{\rm P}_{n}\Xi_{\bm K}(\textstyle\sum_{m=1}^n \delta\bm r_m),\nonumber\\
	\label{eq:Mth}
	{\rm Q}_{n}\Theta^{(n)}_{\bm{K}}{\rm Q}_{n}&={\rm Q}_{n}\Theta^{(n-1)}_{T\bm k}{\rm Q}_{n}
\end{align}
with
\begin{align}
	&\Theta^{(0)}_{T\bm k}=\Theta_{0\bm k},\\
	&{\rm P}_{n}=\vert u_{\bm k n}\rangle\langle u_{\bm k n}\vert+\vert u_{\bm k n+1}\rangle\langle u_{\bm k n+1}\vert,\\
	&({\rm Q}_{n})_{ij}=\delta_{ij}-({\rm P}_{n})_{ij}.
\end{align}
It is straightforward to check that $\Theta^{(N_{\rm occ}-1)}_{T\bm k}$
equals to the identity matrix and that the Pontryagin index for each
$\Theta^{(n)}_{\bm K}$ vanishes.

\bibliography{ref}
\end{document}